\documentclass[twocolumn,showpacs,preprintnumbers,superscriptaddress,amsmath,amssymb,floatfix,letterpaper]{revtex4}%

\usepackage{subfigure}

\usepackage{amssymb}
\usepackage{url}
\usepackage{epsfig}
\usepackage{graphicx}
\usepackage{lineno}
\usepackage{ifpdf}

\usepackage{dcolumn}

\newcolumntype{d}[1]{D{.}{.}{#1}}
\newcolumntype{.}{D{.}{.}{-1}}

\usepackage{ifpdf}

\newif\ifpdf
\ifx\pdfoutput\undefined
   \pdffalse               
\else
   \pdfoutput=1            
   \pdftrue
\fi

\ifpdf
\usepackage[%
  pdftitle={Magnetic moment of 104-Ag-m and hyperfine magnetic field of Ag in Fe using nuclear magnetic resonance
on oriented nuclei},%
  pdfauthor={V.V.~Golovko, et. al.},%
  pdfsubject={Magnetic moments, hyperfine magnetic field, nuclear magnetic resonance
on oriented nuclei},%
  pdfkeywords={nuclear magnetic moment of 104-Ag-m, hyperfine magnetic field
of Ag in Fe, on-line nuclear orientation, nuclear magnetic resonance
on oriented nuclei, configuration mixing, additivity rule},%
  pdfstartview=FitH,%
  bookmarks=true,%
  bookmarksopen=true,%
  breaklinks=true,%
  colorlinks=true,%
  linkcolor=blue,anchorcolor=blue,%
  citecolor=blue,filecolor=blue,%
  menucolor=blue,pagecolor=blue,%
  urlcolor=blue]{hyperref}
\else
\usepackage[%
  breaklinks=true,%
  colorlinks=true,%
  linkcolor=blue,anchorcolor=blue,%
  citecolor=blue,filecolor=blue,%
  menucolor=blue,pagecolor=blue,%
  urlcolor=blue]{hyperref}
\fi

\begin{document}




\title[Short Title]{Magnetic moment of $^{104}$Ag$^{\textit{m}}$ and hyperfine magnetic field of Ag in Fe using nuclear magnetic resonance
on oriented nuclei}


\author{V.V.~Golovko}
\email{vgolovko@owl.phy.queensu.ca}
\altaffiliation{Present address:
Department of Physics, Queen's
University, Stirling Hall, Kingston, ON, Canada, K7L3N6}%
\affiliation{Instituut voor Kern- en Stralingsfysica, Katholieke Universiteit Leuven, Celestijnenlaan 200D, B-3001 Leuven, Belgium}%
\author{I.S.~Kraev}%
\author{T.~Phalet}%
\author{B.~Delaur\'e}%
\author{M.~Beck}%
\author{V.Yu.~Kozlov}%
\author{S.~Coeck}%
\author{F.~Wauters}%
\affiliation{Instituut voor Kern- en Stralingsfysica, Katholieke Universiteit Leuven, Celestijnenlaan 200D, B-3001 Leuven, Belgium}%
\author{P.~Herzog}%
\author{Ch.~Tramm}%
\affiliation{Helmholtz-Institut f\"{u}r Strahlen- und Kernphysik, Universit\"{a}t Bonn, 53115 Bonn, Germany}%
\author{D.~Z\'akouck\'y}%
\author{D.~V\'enos}%
\author{D.~Srnka}%
\author{M.~Honusek}%
\affiliation{Nuclear Physics Institute, ASCR, 250 68 \v{R}e\v{z}, Czech Republic}%
\author{U.~K\"oster}%
\affiliation{Institut Laue Langevin, 6 rue Jules Horowitz, F-38042 Grenoble Cedex 9, France}%
\affiliation{ISOLDE, CERN, 1211 Gen\`eve 23, Switzerland}%
\author{N.~Severijns}%
\affiliation{Instituut voor Kern- en Stralingsfysica, Katholieke Universiteit Leuven, Celestijnenlaan 200D, B-3001 Leuven, Belgium}%


\date{\today}


\begin{abstract}

Nuclear magnetic resonance (NMR/ON) measurements with
$\beta$ and $\gamma$~ray detection have been performed on oriented
$^{104}$Ag$^{\textit{g,m}}$ nuclei with the NICOLE $^3$He-$^4$He dilution
refrigerator setup at ISOLDE/CERN. For $^{104}$Ag$^{\textit{g}}$ ($I^{\pi}=5^+$)
the $\gamma$-NMR/ON resonance signal was found at $\nu = 266.70(5)$~MHz.
Combining this result with the known magnetic moment for this
isotope the magnetic hyperfine field of Ag impurities in an Fe host
at low temperature ($<$ 1 K), is found to be
$|\it{B}_{\rm{hf}}({\rm{Ag}}{\it{Fe}})|$~=~44.709(35)~T. A detailed
analysis of other relevant data available in the literature
yields three more values for this hyperfine field. Averaging all four values
yields a new and precise value for the hyperfine field of Ag
in Fe i.e. $|\it{B}_{\rm{hf}}({\rm{Ag}}{\it{Fe}})|$~=~44.692(30)~T.
For $^{104}$Ag$^{\textit{m}}$ ($I^{\pi}=2^+$), the anisotropy of the
$\beta$~particles provided the NMR/ON resonance signal at
$\nu$~=~627.7(4)~MHz. Using the new value for the hyperfine field of
Ag in Fe this frequency corresponds to the magnetic moment
$\mu(^{104m}\rm{Ag})~ =~+3.691(3)~\mu_{\rm{N}}$, which is
significantly more precise than previous results. The
magnetic moments of the even-A $^{102-110}$Ag isotopes are discussed
in view of the competition between the $(\pi
\rm{g}_{9/2})^{-3}_{7/2^+} (\nu \rm{d}_{5/2} ~ \nu \rm{g}_{7/2})_{5/2^+}$ and
$(\pi \rm{g}_{9/2})^{-3}_{9/2^+} (\nu \rm{d}_{5/2} ~ \nu \rm{g}_{7/2})_{5/2^+}$
configurations. The magnetic moments of the ground and isomeric
states of $^{104}$Ag can be explained by an almost complete mixing
of these two configurations.

\end{abstract}





\maketitle



\section{Introduction}

Nuclear magnetic moments are an important tool for the study of nuclear
structure as they provide a sensitive test of the nuclear coupling
scheme. The light even-A $^{102-110}$Ag isotopes provide a very
interesting case. In the heavier $^{106-110}$Ag isotopes the $(\pi
\rm{g}_{9/2})^{-3}_{7/2^+}$ proton group and the $(\nu
\rm{d}_{5/2})^{-1}_{5/2^+}$ neutron state couple to produce
$I^{\pi}=1^+$ and $I^{\pi}=6^+$ ground and isomeric states,
respectively. In the lighter $^{102-104}$Ag isotopes the
ground and isomeric states have, respectively, $I^{\pi}=5^+$ and
$I^{\pi}=2^+$ and are resulting from a mixture of the $(\pi
\rm{g}_{9/2})^{-3}_{7/2^+}$ and $(\pi \rm{g}_{9/2})^{-3}_{9/2^+}$
proton groups coupling to a $(\nu d_{5/2} ~ \nu g_{7/2})^n_{5/2^+}$
neutron configuration, where $n = 5$ or 7. (Henceforth, for brevity,
the neutron occupation $n$ will be omitted). Thus, whereas in most cases
isomeric states have excitation energies of several hundred keV and a
higher spin than the corresponding ground state, the silver isotopes
$^{102,104}$Ag provide a different picture: the excitation energies
of the isomeric states do not exceed a few keV, and their spins are
lower than those of the ground states. The exact nature of the
mixing of different configurations to produce the wave functions of
these isomeric and ground states turns out to be an intriguing
problem, to which precise values of the nuclear magnetic moments of
these states can provide important information.

The nucleus $^{104}$Ag has a $I^{\pi}=5^+$ ground state with
half-life $t_{1/2}=69$~min, and a $I^{\pi}=2^+$ isomeric state
with $t_{1/2}=33.5$~min at an excitation energy of only 6.9~keV.
For the $I^{\pi}=5^+$ ground state the magnetic moment was
determined with high accuracy in a $\gamma$-NMR/ON measurement as
$\mu =$ 3.914(8) $\mu_{\rm{N}}$ \cite{vandeplassche86} and later also in
an atomic beam magnetic resonance experiment, i.e. $\mu =$
3.919(3)~$\mu_{\rm{N}}$ \cite{dinger89}. For the
isomeric state the only value for the magnetic moment that is
available in the literature, i.e. $\mu
=$ +3.7(2) $\mu_{\rm{N}}$ \cite{ames61}, is not very precise though. Note that
by careful comparison of the works of Ames et al~\cite{ames61},
Greenebaum and Phillips~\cite{greenebaum74}, and
van Walle~\cite{vanwalle85} we found that the value
$\mu=4.12(25)~\mu_{\rm{N}}$ listed for $^{104}$Ag$^{\textit{m}}$
in Table~3.8 of Ref.~\cite{vanwalle85}, and which was copied in Ref.~\cite{raghavan89}
and later also in Ref.~\cite{stone05}, is in fact the value that was obtained by
Greenebaum and Phillips~\cite{greenebaum74} for $^{102}$Ag$^{\textit{m}}$.



Previously already, the proton ground state configuration of
$^{104}$Ag was found to be a mixture of the $(\pi
\rm{g}_{9/2})^{-3}_{7/2^+}$ and $(\pi \rm{g}_{9/2})^{-3}_{9/2^+}$ proton groups
\cite{ames61}. A clear indication for a transition from $(\pi
\rm{g}_{9/2})^{-3}_{7/2^+}$ being the dominant component in the
wavefunction, to $(\pi\rm{g}_{9/2})^{-3}_{9/2^+}$ was found when going from
$^{103}$Ag (with $I^{\pi}=7/2^+$) to $^{101}$Ag (with $I^{\pi}=9/2^+$),
and also when going from
$^{106}$Ag to $^{102}$Ag \cite{vanwalle85}. For the isomeric state
$^{104}$Ag$^{\textit{m}}$ with $I^{\pi}=2^+$ a mixing of these two proton groups
was assumed on the basis of the not very precise magnetic moment value
from an atomic beam experiment \cite{ames61}. We have therefore performed
a new measurement on $^{104}$Ag$^{\textit{m}}$ in order to get a more precise
value for the magnetic moment of this isomeric state, thus
permitting to shed more light on its exact configuration as well as
on the evolution of the proton configuration in the $^{102-106}$Ag
isotopes. The method of nuclear magnetic resonance on oriented
nuclei was used, observing the destruction of the $\beta$ particle
emission asymmetry by radio frequency radiation ($\beta$ NMR/ON). The
nuclei were oriented with the method of low temperature nuclear
orientation \cite{postma86}.

$^{104}$Ag$^{\textit{m}}$ was obtained from the decay of $^{104}$Cd parent nuclei
($t_{1/2}=57.7$~min) that were implanted in the sample foil. As the
$^{104}$Ag ground state was found to be present in the sample as
well, we measured in the same experimental run also the resonance
signal for $^{104g}$Ag. The fact that the magnetic moment of this
state had previously been
determined already with good accuracy from hyperfine structure
measurements, yielding $\mu =$ 3.913(3)~$\mu_{\rm{N}}$ \cite{dinger89},
allowed us to obtain a new and precise value for the hyperfine field
of Ag impurities in Fe.

Note, finally, that the magnetic moment of $^{104m}$Ag determined
here also served as calibration for a measurement of the
$\beta$ emission asymmetry in the decay of this isotope, for the
study of isospin mixing \cite{golovko05b}.

\section{Experimental details}

\subsection{Sample preparation and detection set-up}

The Ag isotopes used for the NMR/ON studies reported here were
obtained from the decay of the $^{104}$Cd precursor produced with
the ISOLDE facility \cite{kugler92,kugler00}. The radioactive $^{104}$Cd
($t_{1/2}$ = 57.7 min) was produced with a 1.4-GeV proton beam
($8\cdot10^{12}$ protons per pulse, staggered mode) from the CERN
Proton Synchrotron Booster accelerator (PS Booster), bombarding a
tin liquid metal target \cite{koster01}. Reaction products
diffused out of the target and effused via a hot transfer line to
the ion source. The atoms were ionized to 1$^+$ ions, extracted,
accelerated to 60 keV and then mass separated by the ISOLDE
General Purpose Separator (GPS). Finally, the $^{104}$Cd beam was
transported through the beam distribution system and implanted
into a 125~$\mu$m thick, 99.99~\% pure iron foil soldered onto the
cold finger of the $^{3}$He-$^{4}$He refrigerator of the NICOLE
low-temperature nuclear orientation set-up
\cite{schlosser88,wouters90}. The iron foil supplied by
Goodfellows$^{\rm{TM}}$ was first polished and then annealed under a
hydrogen atmosphere at $\approx800^{\circ}$C for about 6~hours.

Prior to the implantation of the $^{104}$Cd beam
a polarizing magnetic field of $B_{\rm{ext}}$ = 0.5~T was applied
with the superconducting split coil magnet in the refrigerator, in
order to fully magnetize the iron foil. This field was then
lowered to $B_{\rm{ext}}=$0.1008(3)~T to reduce its influence on
the trajectories of the $\beta$ particles. The fact that part
(i.e. about 10\% \cite{schuurmans96}) of the saturation magnetization
of the Fe foil was lost when the field was reduced to this lower
value slightly reduced the sensitivity but did not further affect
the measurements.

%

\begin{figure}
  \centering
  \includegraphics[width=\columnwidth]{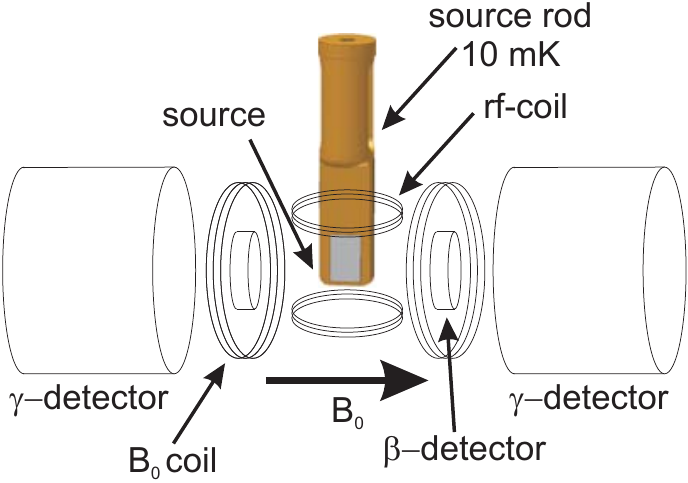}\\
  \caption[Arrangement of RF coils for NMR/ON.]{(Color online) Schematic lay-out of the experimental set-up.
  The radioactive sample is prepared by implanting the ISOLDE beam, that is not shown here and is
  arriving perpendicular to the plane of the drawing, in the Fe foil
  on the Cu sample holder. A polarizing magnetic field $B_{\rm{ext}}$ is created by the split-coil
  superconducting magnet. The RF coil provides a field perpendicular to $B_{\rm{ext}}$ for the NMR/ON
  measurements. The $\beta$ particles
  are observed with planar HPGe detectors installed inside the 4K-shield of the refrigerator at an angle
  of about 15$^\circ$ with respect to the magnetic field $B_{\rm{ext}}$. Large volume HPGe $\gamma$ detectors
  are installed outside the refrigerator. }\label{fig:NMR_view}
\end{figure}

The geometry of the experimental set-up was very similar to the
one we used for our previous $\beta$-NMR/ON experiments
\cite{golovko04,golovko05a,golovko05b} and is shown schematically in
Figure~\ref{fig:NMR_view}.
The angular distribution of beta particles was observed with two planar
HPGe particle detectors \cite{zakoucky04,venos00} with a sensitive
area of about 110~mm$^2$, mounted inside the 4 K radiation shield
of the refrigerator at a distance of about 32~mm from the sample.
Operating these detectors inside the 4K radiation shield, i.e.
without any material between source and detector, avoids energy
losses and reduces scattering of the $\beta$ particles. They were
mounted at an angle of 15$^\circ$ with respect to the magnetization
axis (viz. the horizontal external magnetic field axis) in order
to minimize the influence of scattering effects in the
Fe host foil. Thin isolated copper wires (about 13 cm long)
connected the detectors to the preamplifiers that were placed
outside the refrigerator, resulting in an energy resolution of
about 3~keV for 1~MeV $\beta$ particles.

Angular distributions of $\gamma$ rays were observed with
three large volume HPGe detectors with an energy resolution of
about 3~keV at 1332~keV. These were placed outside the
refrigerator, two along the magnetization axis (see Fig.~\ref{fig:NMR_view})
and one perpendicular to it.
They served to observe the $\gamma$ rays of the Ag isotopes,
as well as to monitor the temperature of the sample
by observing the anisotropy of the 136 keV $\gamma$ ray from a
calibrated $^{57}\rm{Co}\underline{\rm{Fe}}$ nuclear orientation
thermometer \cite{marshak86,kraev05} that was soldered on the back
side of the sample holder.

\subsection{Angular distribution formalism}

The angular distribution of radiation emitted from an axially
symmetric ensemble of oriented nuclei is given by~\cite{krane86}
\begin{linenomath*}
\begin{equation}\label{eq:AngularDistribution}
    W\left( \theta  \right) = 1 + f
    \sum\limits_\lambda  {B_\lambda  \left( {\mu B_{\rm{tot}}/k_B T, I}
    \right) U_\lambda  A_\lambda  Q_\lambda  P_\lambda  \left( {\cos \theta}
    \right)  ,}
\end{equation}
\end{linenomath*}

\noindent where $f$ represents the fraction of the nuclei that
experience the full orienting hyperfine interaction, while the rest $(1-f)$
is supposed to feel no orienting interaction at all; the $B_\lambda$
describe the nuclear orientation; the $U_\lambda$ are the
deorientation coefficients which account for the effect of
unobserved intermediate radiations; the $A_\lambda$ are the
directional distribution coefficients which depend on the
properties of the observed radiation and the nuclear levels
involved; the $Q_\lambda$ are solid angle correction factors and
$P_\lambda( \cos \theta )$ are the Legendre polynomials. The
detection angle $\theta$ is measured relative to the direction of
the saturation magnetization axis of the Fe host foil which is
defined by the applied horizontal magnetic field. The orientation
coefficients $B_\lambda$ depend on the temperature of the sample
$T$, the spin $I$ and the magnetic moment $\mu$ of the oriented
state, and on the total magnetic field $B_{\rm{tot}}$ the nuclei
experience (with $k_B$ the Boltzmann constant). This field is given by

\begin{linenomath*}
\begin{equation}
B_{\rm{tot}}=B_{\rm{hf}}+B_{\rm{ext}}(1+K)-B_{\rm{dem}}
\label{eq:Btot}
\end{equation}
\end{linenomath*}

\noindent where $B_{\rm{hf}}$ is the hyperfine
magnetic field of Ag in Fe (see Sec.IV.A),
$B_{\rm{ext}}$ is the externally applied field, $K$ is the Knight
shift parameter and $B_{\rm{dem}}$ is the demagnetization field \cite{chikazumi64}.

Three measurements of the Knight shift for silver in iron are reported
in the literature. All three determined the Knight shift in
$\gamma$-NMR/ON experiments, yielding $K(^{106}{\rm{Ag}}{Fe})=-0.03(2)$
\cite{eder84}, $K(^{107g}{\rm{Ag}}{Fe})=-0.08(8)$
\cite{eder85} and $K(^{110m}{\rm{Ag}}{Fe})=-0.047(5)$
\cite{duczynski83}. We will use here the weighted average of these three results,
i.e. $K({\rm{Ag}}{Fe})=-0.046(5)$.

For $\gamma$ rays only $\lambda$-even terms occur in
Eq.~(\ref{eq:AngularDistribution}). For positrons from allowed
$\beta$ decays only the $\lambda=1$ term is present and
Eq.(\ref{eq:AngularDistribution}) transforms to~\cite{krane86}

\begin{linenomath*}
\begin{equation}\label{eq:AngularDistributionsElectrons}
    W \left( \theta  \right) = 1 + f \frac{v}{c} B_1 A_1 Q_1 \cos
    \theta ,
\end{equation}
\end{linenomath*}

\noindent where ${v / c}$ is the positron velocity relative to the
speed of light.

Experimentally the angular distribution is obtained by
\begin{linenomath*}
\begin{equation}
W(\theta) = \frac{N_{\rm{cold}}(\theta)}{N_{\rm{warm}}(\theta)} ,
\end{equation}
\end{linenomath*}
\noindent with $N_{\rm{cold,warm}}(\theta)$ the count rates when the
sample is 'cold' (about 10 mK; oriented nuclei) or 'warm' (about 1K;
unoriented nuclei). In on-line experiments, where the count rates
vary with beam intensity, it is customary to construct a double
ratio, combining count rates in two different detectors in order
to eliminate effects of beam intensity fluctuations and avoid the
need to correct for the lifetime of the isotope. In the present
work the double ratio
\begin{linenomath*}
\begin{equation}
R = \left[ \frac{N(15^\circ)}{N(165^\circ)} \right]_{\rm{cold}} /
\left[ \frac{N(15^\circ)}{N(165^\circ)} \right]_{\rm{warm}}
\end{equation}
\end{linenomath*}
\noindent was used for the $\beta$ particles and
\begin{linenomath*}
\begin{equation}
R = \left[ \frac{N(0^\circ)}{N(90^\circ)} \right]_{\rm{cold}} / \left[
\frac{N(0^\circ)}{N(90^\circ)} \right]_{\rm{warm}}
\end{equation}
\end{linenomath*}
\noindent for $\gamma$ rays. All data were corrected for the dead
time of the data acquisition system using a precision pulse
generator.

\subsection{NMR/ON resonance set-up and formalism}

For the NMR/ON measurements an RF oscillating field was applied
perpendicular to the external magnetic field (see
Fig.~\ref{fig:NMR_view}). The NMR coil producing this oscillating
field consisted of a pair of two-turn coils mounted on a teflon
frame that was fixed inside the 300 mK radiation shield surrounding the
sample. The coil was fed from the top of the refrigerator by
coaxial cables connected to a digital Marconi frequency generator with a
range from 10 kHz to 3.3 GHz. In addition to the NMR coil a
pick-up coil for monitoring the RF signal was present too. A
linear RF amplifier with a constant gain of 46~dBm was installed
between the frequency generator and the RF coil. The intensity of
the RF signal was kept as small as possible in order to avoid
too strong RF heating and subsequent reduction in anisotropy for
the $\beta$ particles and $\gamma$ rays.

The resonance measurements were performed at sample temperatures
in the range from 8 to 12~mK by scanning the radio frequency,
$\nu$, and observing the variation of the anisotropy of the
$\beta$ particles and $\gamma$ rays emitted by the Ag nuclei. At resonance,
transitions between the Zeeman split nuclear sublevels are
induced which partially equalize the originally unequal populations of
the sublevels, thus reducing the degree of nuclear orientation and
therefore the magnitude of the anisotropy $R - 1$. The resonance
frequency $\nu_{\rm{res}}$ is related to the nuclear magnetic
moment through the relation
%
%
%
%
\begin{linenomath*}
\begin{equation}
\label{eq:resFreqPractical}
    \nu_{\rm{res}}[\rm{MHz}]
    = \left| \frac { 7.6226  ~  \mu[\mu_{\rm{N}}] ~ \textit{B}_{\rm{tot}}[\rm{T}]  }
                    {\textit{I}[\hbar]}
      \right|
\end{equation}
\end{linenomath*}

%
%
When pro mille precision is required and the nucleus is situated in a medium
other than vacuum and being subject to an external applied magnetic field,
the nuclear magnetic moment value has to be corrected for diamagnetism. This is
a reduction of the field experienced by the nuclei due to the polarization
of the medium, which leads to an apparent reduction in the nuclear magnetic
dipole moment if no correction to the applied field strength is made.
The corrected magnetic moment value is then given by
\begin{linenomath*}
\begin{equation}
\mu_{\rm{corr}} = \mu_{\rm{uncorr}} \left[ 1 - \sigma ~ \frac{B_{\rm{ext}}}{B_{\rm{ext}} + B_{\rm{hf}}} \right]^{-1}
\label{diamagnetic correction}
\end{equation}
\end{linenomath*}
\noindent with $\sigma$ the diamagnetic correction. For silver nuclei
$\sigma$ = 0.005555 \cite{raghavan89}.

\section{Data collection and Analysis}

Due to the rather long half-life of both isotopes and their
rather large magnetic hyperfine interaction strengths $ T_{\rm{int}} =
\left| \mu B /k_{\rm{B}} I \right| $, i.e. $T_{\rm{int}} \simeq 13$~mK for
$^{104}$Ag$^{\textit{g}}$ and $T_{\rm{int}} \simeq 30$~mK for $^{104}$Ag$^{\textit{g}}$,
relaxation of the nuclear spins to thermal equilibrium was
sufficiently fast and no problems with incomplete spin-lattice
relaxation \cite{klein86,shaw89,venos03} were expected.

\begin{figure}
  \includegraphics[width=\columnwidth]{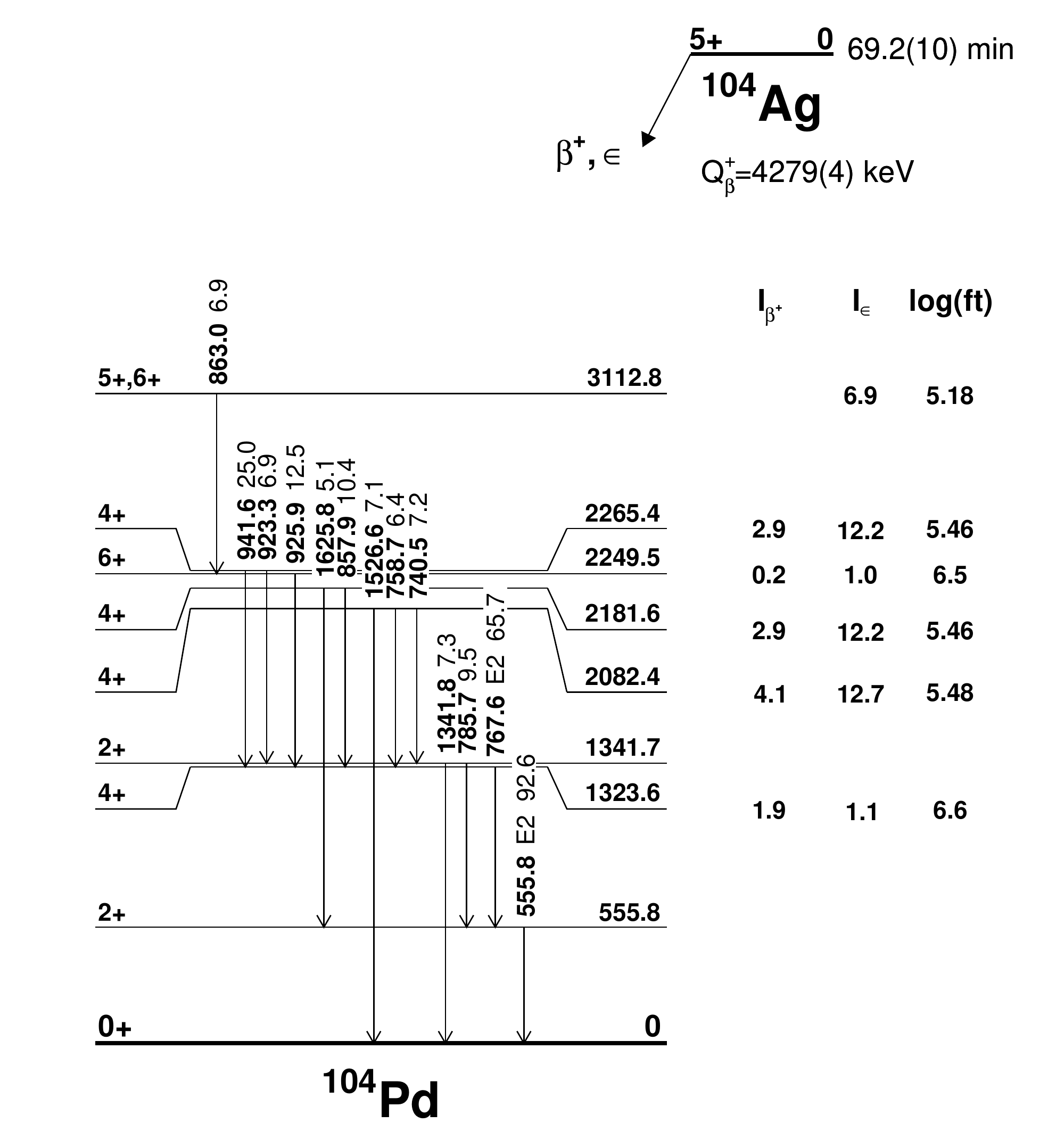}
  \caption[Partial decay scheme of $^{104}$Ag$^{\textit{g}}$.]
  {Partial decay scheme of $^{104}$Ag$^{\textit{g}}$ showing only those
  $\gamma$ transitions with intensities larger than 5\% and the related levels
  and $EC/\beta$ decay branches. The most intense $\gamma$ lines, with energies of
  555.8~keV, 767.6~keV and 941.6~keV, were used for the $\gamma$-NMR/ON
  measurements.}
  \label{fig:104Ag decay scheme}
\end{figure}

The decay of the 5$^+$ $^{104}$Ag$^{\textit{g}}$ ground state
(Figure~\ref{fig:104Ag decay scheme}) is distributed over many
$EC/\beta^+$ branches resulting in three rather intense
$\gamma$ rays with energies of 555.8 keV, 767.6 keV, and 941.6 keV
and intensities of 93\%, 66\% and 25\% , respectively. Whereas the
555.8 keV transition is also clearly present in the decay of the
isomeric state, the $\beta$ decay of the isomer contributes very
little (i.e. only $\simeq 1$\%) to the intensity of the
767.6~keV transition, while the 941.6~keV
transition is even completely absent in the decay of the isomer
\cite{blachot07}. The last two $\gamma$ rays are therefore
especially well suited for the determination of the magnetic
moment of $^{104}$Ag$^{\textit{g}}$ in a $\gamma$-NMR/ON experiment.


For $^{104}$Ag$^{\textit{m}}$ the main $EC/\beta^+$ branch represents
$\simeq 70$\% of the total decay strength and populates the 2$^+$
level at 555.8~keV in $^{104}$Pd \cite{blachot07}. All other
branches represent less than 2\% each of the total decay strength.
As the 555.8~keV $\gamma$ transition depopulating the 2$^+$ first
excited state is also one of the stronger $\gamma$ lines in the decay
of $^{104}$Ag$^{\textit{g}}$ (see Fig.~\ref{fig:104Ag decay scheme})
and the next most intense $\gamma$ transition in the decay of
$^{104}$Ag$^{\textit{m}}$ has an intensity of only about 4\%,
it was decided to use the $\beta$ transition to the first excited
state to search for the resonance signal of $^{104m}$Ag.

\begin{figure}
  \includegraphics[width=\columnwidth]{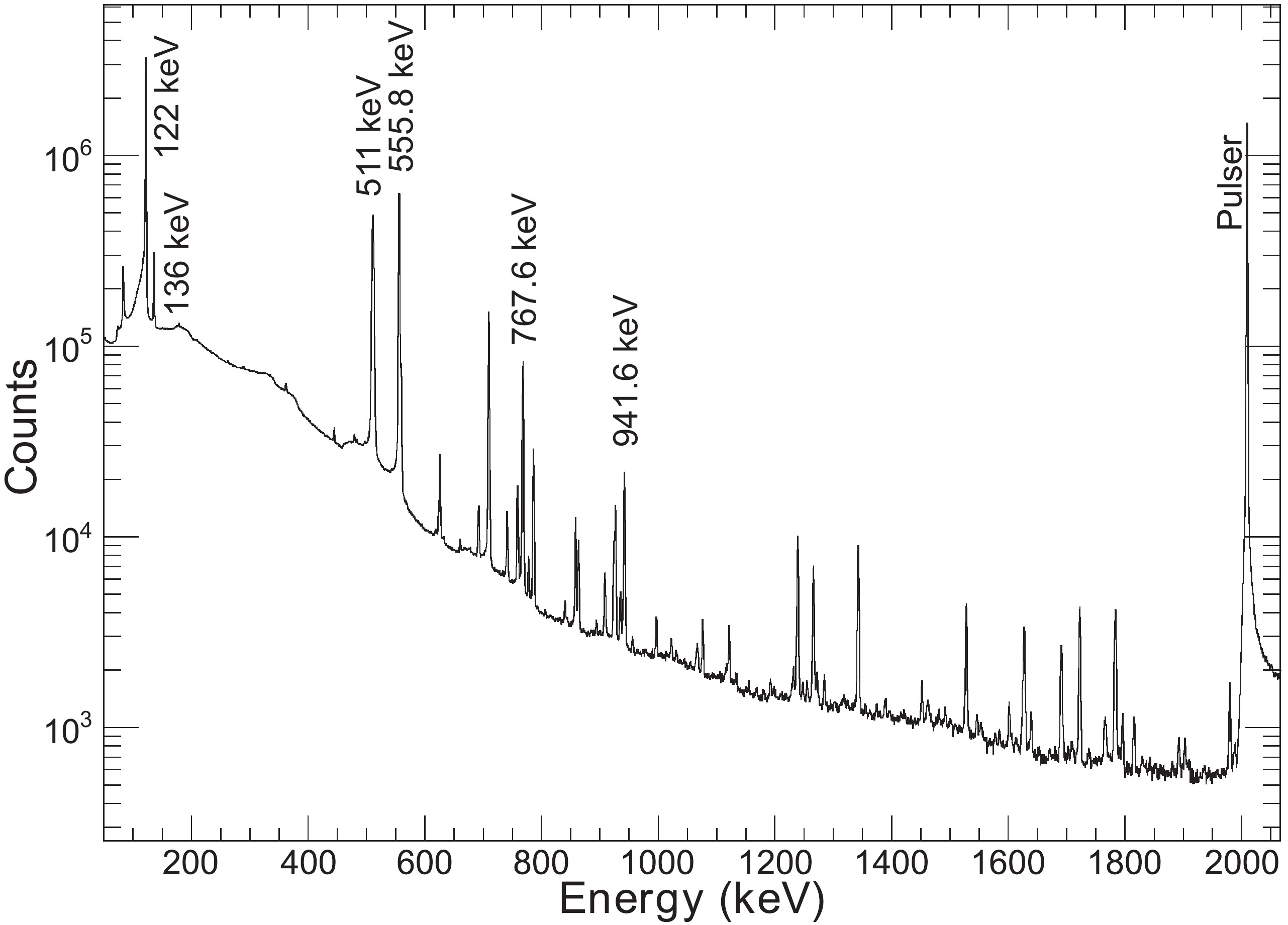}
  \caption[Typical $\gamma$ spectrum for $^{104}$Ag$^{\textit{g,m}}$.]
  {Typical $\gamma$ spectrum for $^{104}$Ag$^{\textit{g,m}}$, obtained in 300 s of counting. The 511~keV
  positron annihilation line, the 122~keV and 136~keV $\gamma$ lines of the $^{57}{\rm{Co}}{Fe}$
  nuclear thermometer, and the 555.8~keV, 767.6~keV, 941.6~keV $\gamma$ ray
  lines that were used for the $\gamma$-NMR/ON measurements on $^{104g}$Ag, are indicated. Most
  $\gamma$ lines belong to the decay of $^{104g,m}$Ag
  (see \cite{blachot07}). The strong line between the 555.8~keV
  and the 767.6~keV lines is the 709~keV internal transition in $^{104}$Cd. }
  \label{fig:gammaSpec104gAg}
\end{figure}

\begin{figure}
  \includegraphics[width=\columnwidth]{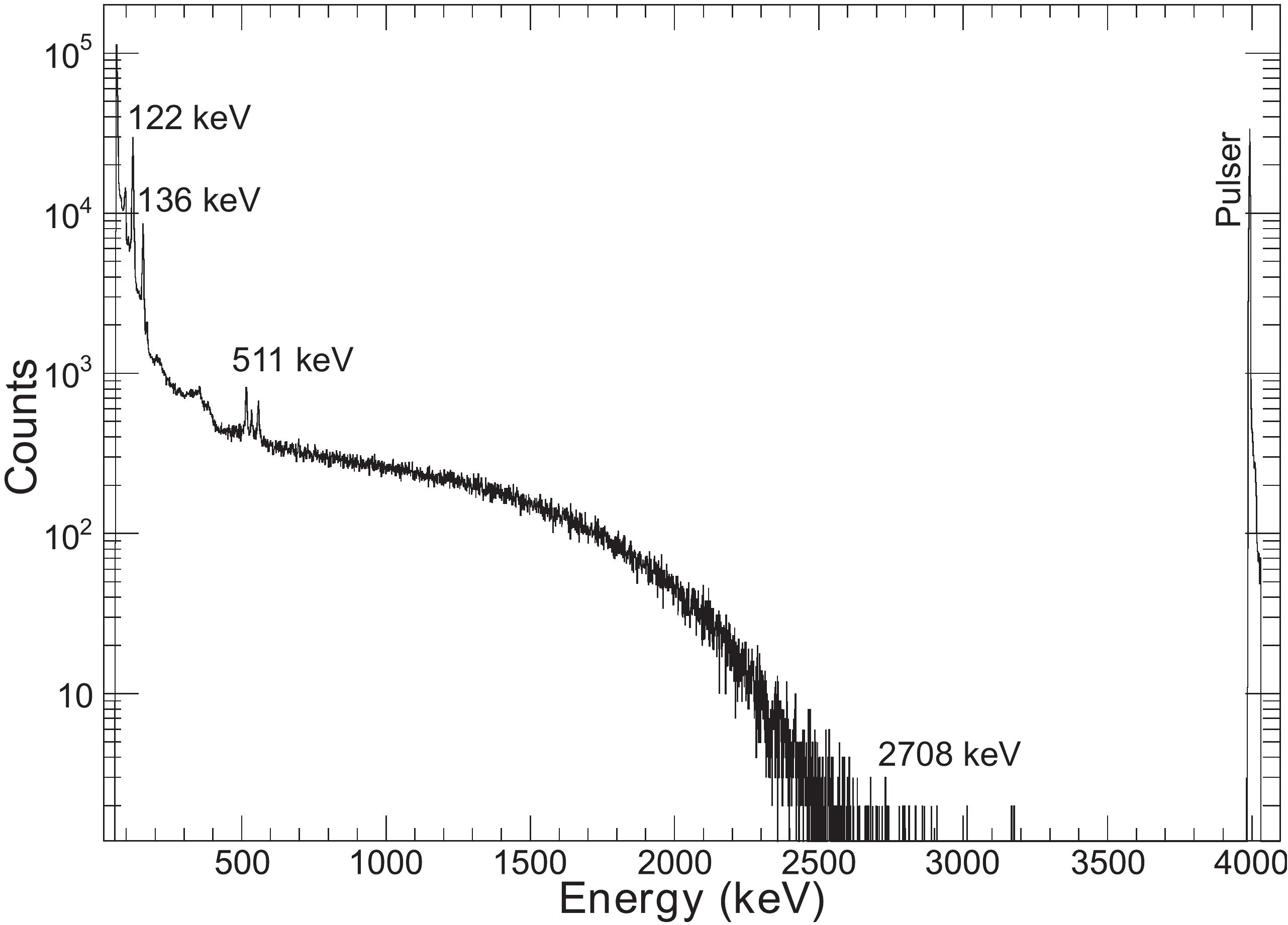}
  \caption[Typical $\beta$ spectrum for $^{104}$Ag$^{\textit{m}}$.]{Typical
  $\beta$ spectrum for $^{104}$Ag$^{\textit{m}}$ , obtained in 300 s of counting. The endpoint energy for
  the decay of $^{104}$Ag$^{\textit{m}}$ (i.e. 2708~keV) is indicated. At low energy the
  122~keV and 136~keV $\gamma$ rays from the $^{57}{\rm{Co}}{Fe}$ nuclear thermometer, the 511~keV annihilation
  peak and the 555.8~keV $\gamma$ ray from the $^{104}$Ag$^{\textit{g,m}}$ decays are also visible.}
  \label{fig:Spec104mAgBetGr2200}
\end{figure}

Typical $\gamma$ ray and $\beta$ particle spectra are shown in
Figs.~\ref{fig:gammaSpec104gAg} and~\ref{fig:Spec104mAgBetGr2200},
respectively.

\subsection{$\gamma$-NMR/ON on $^{104g}{\rm{Ag}}{Fe}$}

The nuclear magnetic resonance signal for the ground state $^{104}$Ag$^{\textit{g}}$
was obtained by observing the change of the anisotropy of the 555.8,
767.6, 941.6~keV $\gamma$ rays as a function of the frequency of the
applied RF field. Using a triangular modulation frequency of 100 Hz,
a modulation bandwith of 0.5 MHz and an RF signal level of -32
dBm, the center RF frequency was varied in steps of 0.5~MHz over
the resonance search region from 263.5~MHz to 270~MHz.
The search region could be chosen so narrow since two precise and
mutually consistent values are available in the literature for both
the hyperfine field of Ag in Fe, i.e. $|\it{B}_{\rm{hf}}$(Ag$Fe$)$|$~=~44.7~T
\cite{fox71,eder84}, and the magnetic moment of $^{104}$Ag$^{\textit{g}}$, i.e.
$\mu = 3.92 \mu_{\rm{N}}$ \cite{vandeplassche86,dinger89}. Two scans
were performed, stepping the frequency region in both upward and
downward directions, in order to avoid possible shifts of the
(effective) resonance centers due to a finite spin-lattice
relaxation time. At each frequency data were accumulated for 300 s.

\begin{figure*}
    \subfigure[]{
    \label{fig:mini:subfig:a} 
    \begin{minipage}[b]{0.48\linewidth}
        \centering
        \includegraphics[width=1.0\textwidth]{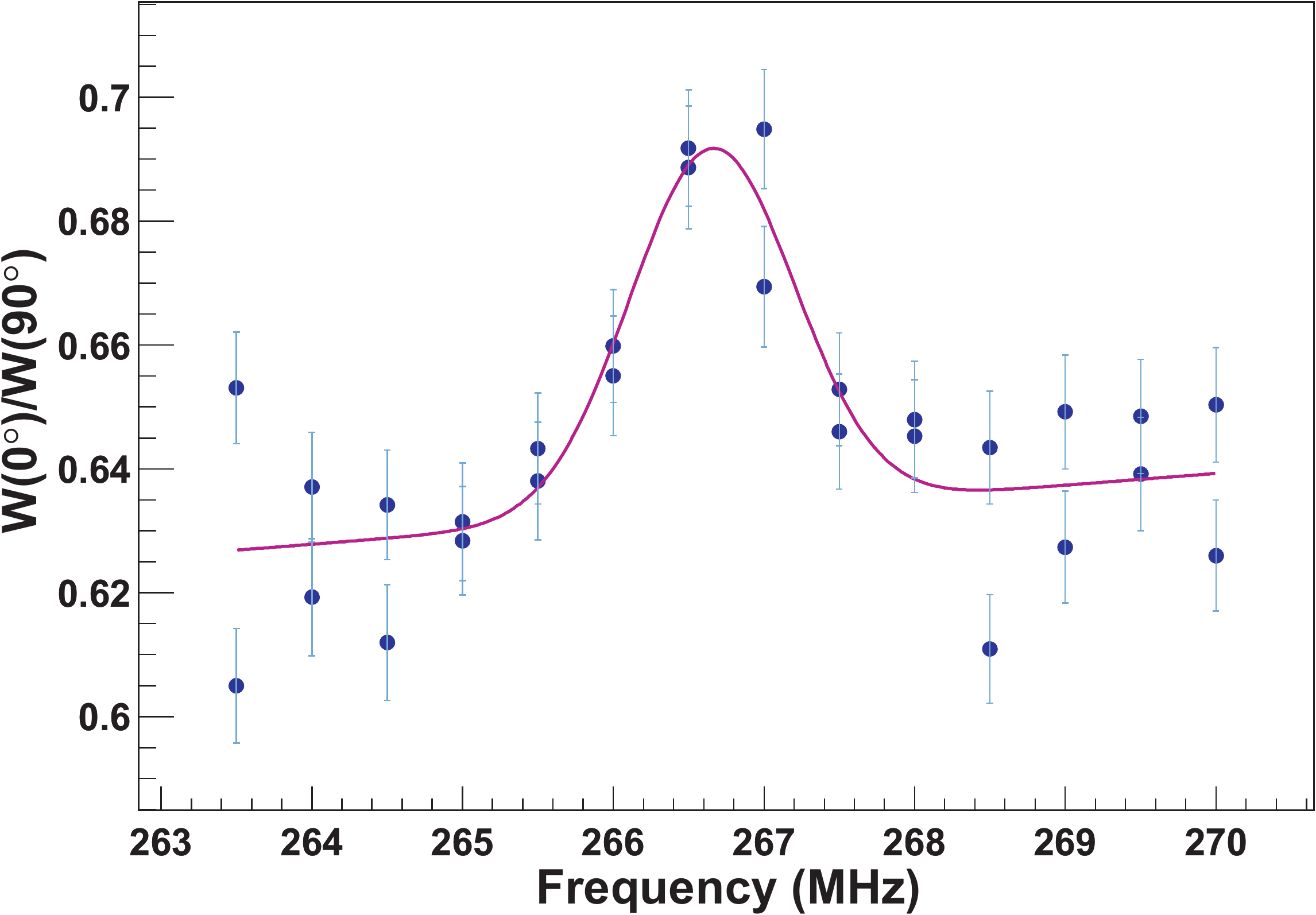}
    \end{minipage}}%
    \hspace{0.1in}
    \subfigure[]{
    \label{fig:mini:subfig:b} 
    \begin{minipage}[b]{0.48\linewidth}
        \centering
        \includegraphics[width=1.0\textwidth]{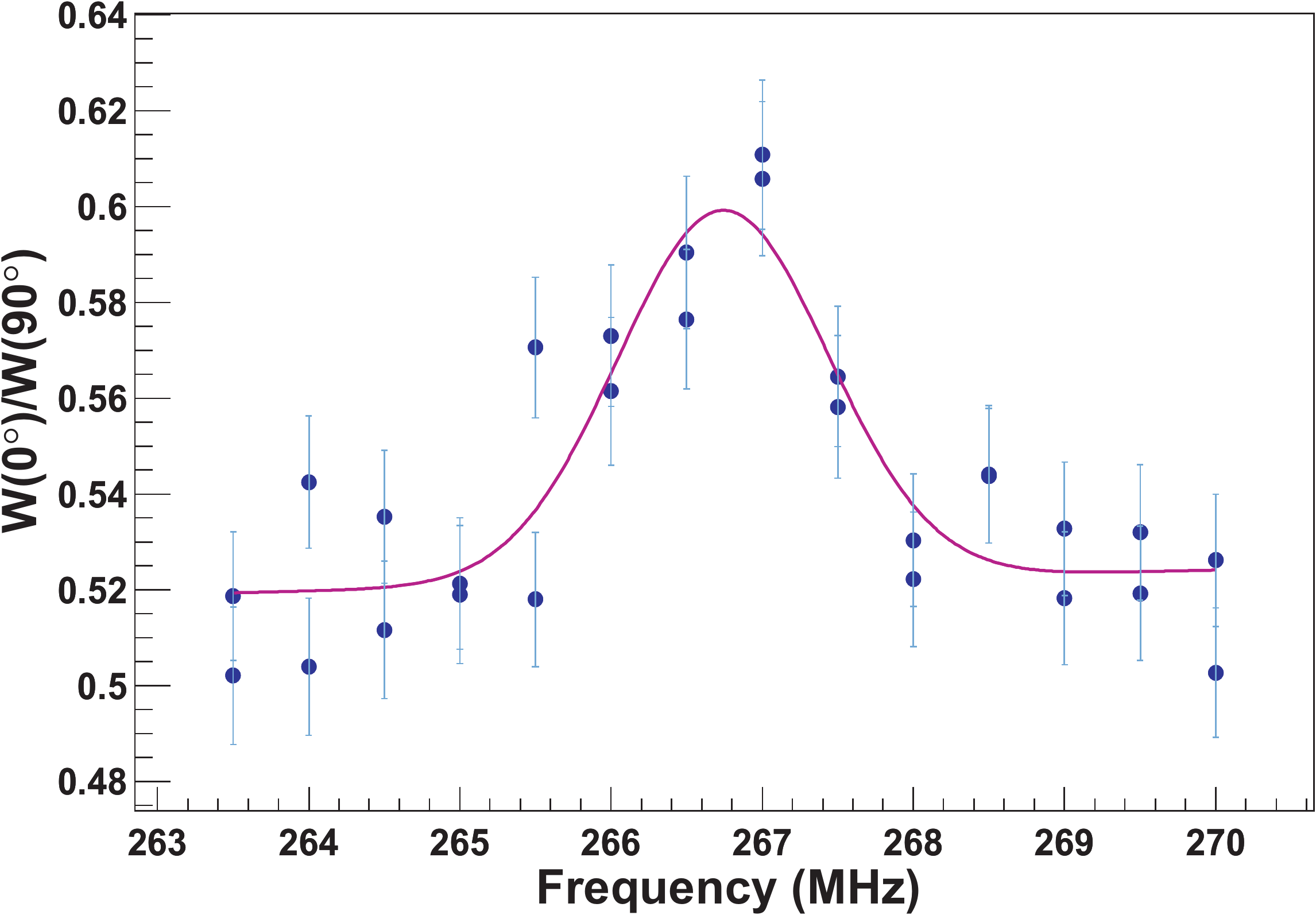}
    \end{minipage}}%
    \caption{(Color online) On-line $\gamma$-NMR/ON curves for the 767.6~keV
        (a) and the 941.6~keV (b) $\gamma$ lines
        of  $^{104}$Ag$^{\textit{g}}$. The ratio of the pulser-normalized
        $\gamma$ anisotropies $W(0^{\circ})/W(90^{\circ})$
        is plotted as a function of RF frequency. Data
        points for two scans (one in upward and one in downward direction)
        are superposed. The integrated destruction of
        $\gamma$ anisotropy is about 25\% in both cases. The corresponding
        resonance frequencies are given
        in Table~\ref{tab:NMRfrequencyAg104g}.}
    \label{fig:mini:subfig} 
\end{figure*}

An NMR effect was observed for all three $\gamma$ rays
(see e.g. Fig.~\ref{fig:mini:subfig}). Table~\ref{tab:NMRfrequencyAg104g}
summarizes the results of the fits of a Gaussian line and a linear
background to the data for the three different lines
with the MINUIT minimization package \cite{james75}.
The weighted average of the three central frequencies, obtained in a
magnetic field of $B_{\rm{ext}}$ = 0.1008(3)~T, is
\begin{linenomath*}
\begin{equation}
|\nu_{\rm{res}}(^{104}{\rm{Ag}}^{\textit{m}}{Fe})| = 266.70(5)~\rm{MHz} .
\end{equation}
\end{linenomath*}
\noindent This value is in agreement with, but an order of magnitude
more precise than the
previous result obtained in a similar experiment by Vandeplassche et al.
\cite{vandeplassche86}, i.e. $\nu_{\rm{res}}=266.3(5)$~MHz. It was used to
determine a new precision value for the hyperfine field of Ag in Fe,
as will be discussed below.


%
%
%

%
%
\begin{table}
\begin{center}
\caption[Overview of the $\gamma$-NMR/ON resonance frequencies
($\nu_{\rm{res}}$) obtained for the three most intensive
$\gamma$ rays of $^{104}$Ag$^{\textit{g}}$.] {$\gamma$-NMR/ON resonance centre
frequencies ($\nu_{\rm{res}}$) obtained for the three
most intense $\gamma$ rays in the decay of $^{104}$Ag$^{\textit{g}}$.}
\label{tab:NMRfrequencyAg104g}
\begin{ruledtabular}
\begin{tabular}{c d}
  Energy (keV) & \multicolumn{1}{c}{\mbox{\hspace*{4mm} $\nu_{\rm{res}}$ (MHz)}} \\
       \\[-3mm]
  \hline
       \\[-3mm]
  555.8 & 266.77(13) \\
  767.6 & 266.66(6) \\
  941.6 & 266.74(9) \\
  \cline{2-2}
  \hline
       \\[-3mm]
  weighted average  & 266.70(5) \\
\end{tabular}
\end{ruledtabular}
\end{center}
\end{table}

\subsection{$\beta$-NMR/ON on $^{104}\rm{Ag}^{\textit{m}}\rm{Fe}$}

To determine the magnetic moment of the isomeric state
the change of the $\beta$ particle anisotropy as a function of the
RF frequency was observed. The only value for the magnetic moment
of $^{104}$Ag$^{\textit{m}}$ available in the literature is not very precise,
i.e. $\mu(^{104}\rm{Ag}^{\textit{m}}$) = +3.7(2)$\, \mu_{\rm{N}}$ \cite{ames61}.
The magnetic moment was therefore first determined, prior to our
measurement but during the same beam time, by scanning the
frequency of the first of the two lasers used to selectively
ionize Ag atoms in the RILIS ion source~\cite{fedosseev03}.
On-line analysis of this laser scan yielded
\begin{linenomath*}
\begin{equation}
\mu(^{104}\rm{Ag}^{\textit{m}}) = 3.7(1) ~ \mu_{\rm{N}}    ,
\end{equation}
\end{linenomath*}
corresponding to a frequency $\nu_{\rm{res}}=630 \pm 17$~MHz,
which was then used as the search region for the $\beta$-NMR/ON
experiment.

In order to maximize statistics, the destruction of the
$\beta$ anisotropy effect
\begin{linenomath*}
\begin{equation}
R=1- \frac {W(15^{\circ})} {W(165^{\circ})}
\end{equation}
\end{linenomath*}
\noindent was monitored in the energy region from 600~keV to the
endpoint at 2708~keV. In spite of the fact that the magnitude of a
$\beta$ anisotropy decreases towards lower energies because of its
dependence on $v/c$ (see
Eq.~(\ref{eq:AngularDistributionsElectrons})), the size of the
anisotropy was still rather large, i.e. $R \simeq 0.50$.
The energy region below 600~keV was
not used as it suffered from background of Compton scattered
511~keV annihilation and 555.8~keV decay $\gamma$ rays.
A triangular modulation frequency of 100~Hz was again used.
Several frequency scans were performed. For most of these the
frequency step width was 2~MHz and the modulation amplitude
$\pm$1~MHz with a nominal RF power level of $-42$~dBm. For each scan
the frequency region was always stepped both in upward and in
downward directions. No difference between the center frequencies
was found for passes in opposite directions, indicating that
relaxation effects were indeed negligible.

\begin{figure}
  \includegraphics[width=\columnwidth]{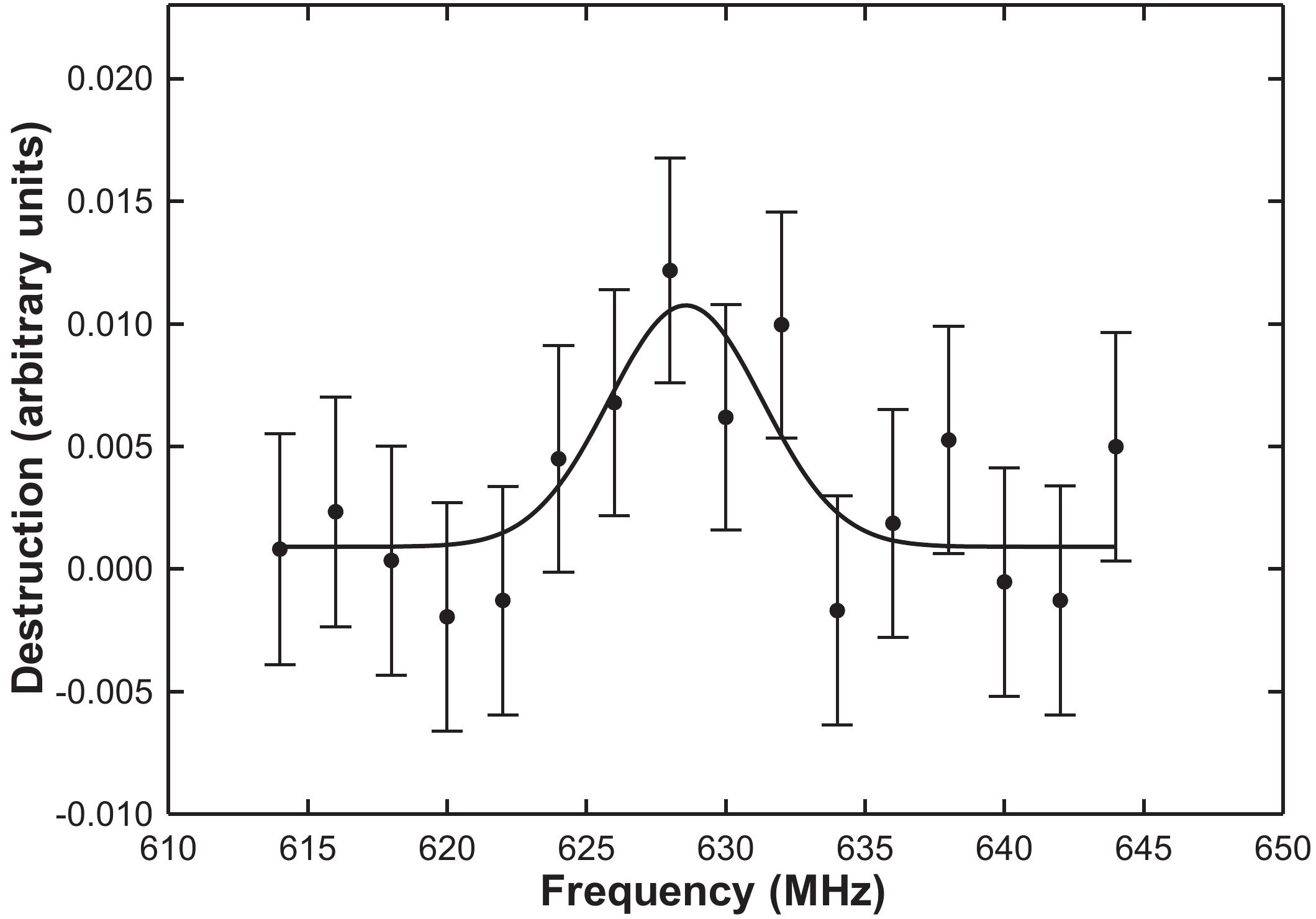}
  \caption[$\beta$-NMR/ON of $^{104}{\rm{Ag}}^{\textit{m}}{Fe}$ ~at applied
  field of $B_{\rm{ext}}=0.1008(3)$~T.]
  {$\beta$-NMR/ON resonance curve for $^{104}{\rm{Ag}}^{\textit{m}}{Fe}$. Data from three frequency scans using
  the FM-off/FM-on sequence have been summed. The destruction of anisotropy is defined in
  Eq.~(\ref{eq:destruction}). The total destruction observed
  was about 4~\%.}
  \label{fig:betaNMR104mAg}
\end{figure}

The first scans were performed with continuous modulation of the
RF signal. These data were again fitted using a simple Gaussian
with a constant background.
Thereafter four scans were performed by recording the
spectrum for each frequency first without and immediately
thereafter with frequency modulation. This FM-off/FM-on mode was
used to obtain a better definition of the line shape. This was
considered to be helpful since resonance lines in iron hosts at
high frequencies are rather broad due to the inhomogeneous
broadening of about 1\%. Fig.~\ref{fig:betaNMR104mAg} shows the
destruction, $S$, of the $\beta$ asymmetry as a function of
frequency. This destruction $S$ is defined as the difference between
the ratio $W(15^{\circ})/W(165^{\circ})$ with FM~on and FM~off
normalized to the ratio with FM~off:
\begin{linenomath*}
\begin{equation}\label{eq:destruction}
    S = \frac
        {\left(
          \left[ \frac {W(15^{\circ})} {W(165^{\circ})} \right]_{\rm{FM~on}} -
          \left[ \frac {W(15^{\circ})} {W(165^{\circ})} \right]_{\rm{FM~off}}
        \right)}
        {\left[ \frac {W(15^{\circ})} {W(165^{\circ})} \right]_{\rm{FM~off}} }
\end{equation}
\end{linenomath*}
The total destruction observed was about 4~\%. This value is smaller
than in our previous experiments with $^{59}$Cu ($S$~$\approx$~45\%)
\cite{golovko04} and $^{69}$As ($S$~$\approx$~10\%) \cite{golovko05a}
and also smaller than the destruction obtained here for the $^{104}$Ag$^{\textit{g}}$
resonances (i.e. $S$~$\approx$~25\%) (Figure~\ref{fig:mini:subfig}).
This is most probably due to the non-resonant background from $\beta$
particles of the non-resonant $^{104g}{\rm{Ag}}{Fe}$ in the energy
region used for analysis.
Nevertheless, the resonance frequency value could still be determined with
good accuracy. Table~\ref{tab:betaNMRfreqAg104m} lists the results
for the various scans that were performed. The weighted average
resonance frequency, again obtained in a magnetic field of
$B_{\rm{ext}}$ = 0.1008(3)~T is
\begin{linenomath*}
\begin{equation}
\nu_{\rm{res}}(^{104}{\rm{Ag}}^{\textit{m}}{Fe}) = 627.7(4)~{\rm{MHz}} .
\end{equation}
\end{linenomath*}

%
%

\begin{table}
\begin{center}
\caption[Overview of the $\beta$-NMR/ON results for various
NMR/ON scans of $^{104}$Ag$^{\textit{m}}$.]{$\beta$-NMR/ON results
for various scans of $^{104}$$^{\textit{m}}$Ag. Every measurement consists
of two scans over the frequency region from 613~MHz to 647~MHz, one
in upward and the other in downward direction. The measurements were
performed with different stepsize and for different measurement times.
The weighted average of the seven individual measurements (i.e. 14 scans)
is indicated as well.}
\label{tab:betaNMRfreqAg104m} 
\begin{ruledtabular}
\begin{tabular*}{\textwidth}{@{\extracolsep{\fill}}cclcc}
  Meas. & Modulation & $\nu_{\rm{res}}$ (MHz) & Step (MHz)  &  Coll. time (s)\\
  \hline
       \\[-3mm]
  1 & on     & 628.3(8)                   & 2   &  150\\
  2 & on     & 627.0(8)                   & 2   &  300\\
  3 & on     & 628.3(9)                   & 5   &  300\\
  4 & off/on & 627.2(7)                   & 2   &  300\\
  5 & off/on & 628.0(11)                  & 2   &  150\\
  6 & off/on & 629.1(29)                  & 2   &  150\\
  7 & off/on & 628.1(19)                  & 2   &  150\\
  \cline{3-3}
  \hline
       \\[-3mm]
  $\overline{\nu}_{\rm{res}}$  &        &   627.7(4)         &  \\
\end{tabular*}
\end{ruledtabular}
\end{center}
\end{table}

\section{Results}

\subsection{Hyperfine field of Ag impurities in Fe}

To extract a precise magnetic moment value for $^{104}$Ag$^{\textit{m}}$ from the
NMR/ON frequency obtained, a precise value for the hyperfine
magnetic field of Ag impurities in Fe host,
$\it{B}_{\rm{hf}}$(Ag$Fe$), is required. Two values are quoted in
the literature. The first,
$|\it{B}_{\rm{hf}}$(Ag$Fe$)$|$~=~44.72(2)~T, was obtained in a
NMR/ON experiment with $^{110}$Ag$^{\textit{m}}$ in Fe \cite{fox71}. However, from
the reported resonance frequency of 203.75(10)~MHz in a field of
0.220~T, and with the magnetic moment of 3.604(4)~$\mu_{\rm{N}}$
used by the authors, one obtains
$|\it{B}_{\rm{hf}}$(Ag$Fe$)$|$~=~44.72(5)~T, i.e. the same value but with
a significantly larger error bar.
It seems that the error on the magnetic moment was
neglected in Ref.~\cite{fox71}. Later, Eder et al.~\cite{eder84}
quoted the value $|\it{B}_{\rm{hf}}$(Ag$Fe$)$|$~=~44.69(5)~T which
they deduced from the NMR/ON frequency
$\nu(^{110}{\rm{Ag}}^{\textit{m}}{Fe})$~=~204.78(2)~MHz reported in
Ref.~\cite{ruter83} and the magnetic moment value
$|\mu(^{110}{\rm{Ag}}^{\textit{m}})|$~=~3.607(4)~$\mu_{\rm{N}}$ from
Refs.~\cite{fischer75} and \cite{feiock69}. In the mean time several
new hyperfine interaction measurements on $^{110}$Ag$^{\textit{m}}$, $^{106}$Ag$^{\textit{m}}$
and $^{104}$Ag$^{\textit{g}}$ have been reported, allowing one to deduce four
precise values for $\it{B}_{\rm{hf}}$(Ag$Fe$), as will be discussed
in the following paragraphs. An overview is given in
Table~\ref{table:hyperfine field AgFe}.

Two values for $\it{B}_{\rm{hf}}$(Ag$Fe$) can be obtained from
data for $^{110}$Ag$^{\textit{m}}$. The two magnetic moment
values for $^{110}$Ag$^{\textit{m}}$ reported in Refs.~\cite{schmelling67}
and \cite{hutchison92} agree well and, after being corrected
for diamagnetism \cite{raghavan89}, lead to the weighted average
value $|\mu(^{110}{\rm{Ag}}^{\textit{m}})|$~=~3.608(3)~$\mu_{\rm{N}}$.
%
%
Combining this with the resonance frequency for this isotope in Fe
host reported in Ref.~\cite{ruter83}, i.e.
$\nu(^{110}{\rm{Ag}}^{\textit{m}}{Fe})$~=~204.78(2)~MHz,
Eq.~(\ref{eq:resFreqPractical}) yields for the hyperfine field of Ag
impurities in Fe the value
\begin{linenomath*}
\begin{equation}
|{\it{B}_{\rm{hf}}}({\rm{Ag}}Fe)|_1~=~44.675(37)~\rm{T} .
\end{equation}
\end{linenomath*}
\noindent Note that no correction for the Knight shift is required
as the resonance frequency was quoted for $B_{\rm{ext}}$ = 0, while
a correction for $B_{\rm{dem}}$ can be neglected at the present level
of precision since a very thin foil was used.

Further, the nuclear magnetic moment of $^{110}$Ag$^{\textit{m}}$ cited above, i.e.
$|\mu(^{110}{\rm{Ag}}^{\textit{m}})|$~=3.608(3)~$\mu_{\rm{N}}$, and
the resonance frequency
$|\nu(^{110m}{\rm{Ag}}{Fe})|$~=~203.75(10)~MHz reported in
Ref.~\cite{fox71}, yield
$|\it{B}_{\rm{tot}}$(Ag$Fe$)$|$~=~44.451(43)~T. The frequency was
obtained with a 3~$\mu$m thin foil, rendering the correction for
$B_{\rm{dem}}$ again negligible, and in $B_{\rm{ext}}$ = 0.220 T.
Taking then also the Knight shift parameter $K$~=~-0.046(5) into
account (which was not yet known at the time of Ref.~\cite{fox71})
one derives from Eq.~(\ref{eq:Btot})
\begin{linenomath*}
\begin{equation}
|{\it{B}_{\rm{hf}}}({\rm{Ag}}Fe)|_2~=~44.661(43)~\rm{T} .
\end{equation}
\end{linenomath*}
%


A third value for $|\it{B}_{\rm{hf}}$(Ag$Fe$)$|$ is obtained from
data for $^{106m}$Ag:
%
%
NMR/ON resonance frequencies for this isotope were determined in both
a Ag and an Fe host. Eder et al.~\cite{eder85}
obtained $|\nu(^{106}{\rm{Ag}}^{\textit{m}}{Fe})|$~=~210.57(3)~MHz in $B_{\rm{ext}}$ = 0 with
a stack of foils of 2~$\mu$m thickness (such that $B_{\rm{dem}}$ = 0). Ohya et al.
\cite{ohya01} reported $|\nu(^{106}{\rm{Ag}}^{\textit{m}}{Ag})|$~=~56.128(26)~MHz
in an external field of about 12~T. The exact value of this applied field
could be deduced from the resonance frequency of $^{110}$Ag$^{\textit{m}}$ in Ag for the same
field setting, which was found at 54.640(1)~MHz \cite{ohya01b}.

%
Using for the magnetic moment of $^{110m}$Ag the weighted average value
from Refs. \cite{schmelling67} and \cite{hutchison92}, i.e.
$|\mu(^{110}{\rm{Ag}}^{\textit{m}})_{\rm{uncorr}}|$=3.588(3)$\mu_{\rm{N}}$
(this time not corrected for diamagnetism as the nuclear orientation in the NMR/ON
experiments of Refs.~\cite{ohya01} and \cite{ohya01b} was induced solely
by an external magnetic field), Eq.~(\ref{eq:resFreqPractical})
yields $|B_{\rm{tot}}|$=11.987(10)\rm{T}.
The ratio $r$~=~3.7516(18) of the two above mentioned frequencies for $^{106}$Ag$^{\textit{m}}$
can be written as
\begin{linenomath*}
\begin{eqnarray}
r & = & \frac{\mu(^{106}{\rm{Ag}}^{\textit{m}})_{\rm{corr}} ~ ~
|\it{B}_{\rm{hf}}({\rm{Ag}}{\it{Fe}})|}{\mu(^{106}{\rm{Ag}}^{\textit{m}})_{\rm{uncorr}}
~ ~  |B_{\rm{tot}}| }
  \nonumber \\
           & = & \frac{ 1.005586 ~ ~ |\it{B}_{\rm{hf}}({\rm{Ag}}{\it{Fe}})| }{11.987(10) ~ \rm{T} }   ,
\label{eq:ratio_106mAg}
\end{eqnarray}
\end{linenomath*}
\noindent with $\mu(^{106}{\rm{Ag}}^{\textit{m}})_{\rm{corr}} / \mu(^{106}{\rm{Ag}}^{\textit{m}})_{\rm{uncorr}}$ = 1.005586
the factor for the diamagnetic correction \cite{raghavan89}, so that
\begin{linenomath*}
\begin{equation}
|{\it{B}_{\rm{hf}}}({\rm{Ag}}Fe)|_3~=~44.720(43)~\rm{T} .
\end{equation}
\end{linenomath*}

Finally, a hyperfine field value can also be obtained from data for
$^{104}$Ag$^{\textit{g}}$. Indeed, combining our NMR/ON resonance frequency for
$^{104}$Ag$^{\textit{g}}$ in Fe, i.e. $|\nu(^{104}{\rm{Ag}}^{\textit{m}}Fe)|$~=~266.70(5)~MHz,
with the magnetic moment value
$|\mu(^{104}{\rm{Ag}}^{\textit{m}})|$~=~3.919(3)~$\mu_{\rm{N}}$ from optical
hyperfine spectroscopy measurements \cite{dinger89},
Eq.~(\ref{eq:resFreqPractical}) yields
$|\it{B}_{\rm{tot}}$(Ag$Fe$)$|$~=~44.639(35)~T. Correcting this
result according to Eq.~(\ref{eq:Btot}) for the external magnetic
field $B_{\rm{ext}}$=0.1008(3)~T, for the Knight shift parameter
$K$~=~$-$0.046(5), and for the demagnetization of the 125~$\mu$m
thick Fe foil used, i.e. $B_{\rm{dem}}$~=~0.026(5)~T (a 20\% error
was adopted in order to account for approximations made in the
equations used to calculate $B_{\rm{dem}}$ \cite{golovko05b}),
results in
\begin{linenomath*}
\begin{equation}
|{\it{B}_{\rm{hf}}}({\rm{Ag}}Fe)|_4~=~44.709(35)~\rm{T} .
\end{equation}
\end{linenomath*}
%


When combining the four above mentioned hyperfine fields to one weighted average,
the correlations between the first three values, which all rely on the same value
for the magnetic moment of $^{110m}$Ag, were duly taken into account by
incorporating the full covariance matrix (e.g.~\cite{barlow89}). Using then further
standard error propagation techniques yields for the magnetic hyperfine field
of Ag impurities in Fe the value
\begin{linenomath*}
\begin{equation}
|{\it{B}_{\rm{hf}}}({\rm{Ag}}Fe)|~=~44.692(30)~\rm{T}
\end{equation}
\end{linenomath*}
\noindent (the error would be 0.020~T instead of 0.030~T if correlations
would not have been taken into account). This value is in agreement with
and more precise than the value of 44.69(5)~T cited in \cite{eder84}. It will
be used further in the analysis of our NMR/ON results for $^{104}$Ag$^{\textit{m}}$
in the next section.

\begin{table*}
\begin{center}
\caption{Input data leading to the hyperfine magnetic field for Ag
impurities in Fe and the values obtained from these.}
\label{table:hyperfine field AgFe}
\begin{ruledtabular}
\begin{tabular}{cc d cl}
  line  &
         \begin{tabular}{c}
           measured/deduced \\
           quantity \\
         \end{tabular}
        &
            \multicolumn{1}{c}{\mbox{\hspace*{10mm} value}}
        &
          \begin{tabular}{c}
            \mbox{Ref.} \\
          \end{tabular}
        &
          \begin{tabular}{c}
            \mbox{remark} \\
          \end{tabular}
           \\
   \hline
       \\[-3mm]
  1  &    $|\mu(^{110}{\rm{Ag}}^{\textit{m}})|$                 &  3.607(4) \mbox{~$\mu_{\rm{N}}$\footnotemark[1]} &  \cite{schmelling67}  &   ABMR~\footnotemark[2]  \\
  2  &    $|\mu(^{110}{\rm{Ag}}^{\textit{m}})|$                 &  3.609(4) \mbox{~$\mu_{\rm{N}}$\footnotemark[1]} &  \cite{hutchison92}   &  BF-NMR/ON~\footnotemark[3]  \\
  3  &    $|\mu(^{110}{\rm{Ag}}^{\textit{m}})|$                 &  3.608(3) \mbox{~$\mu_{\rm{N}}$}                  &                       &  weighted average of above two values  \\
  4  &    $|\nu(^{110}{\rm{Ag}}^{\textit{m}}Fe)|$               &  204.78(2) \mbox{~MHz}                            &  \cite{ruter83}       &  NMR/ON, for $B_{\rm{ext}}$~=~0~T          \\
  5  &    $|\it{B}_{\rm{hf}}({\rm{Ag}}{\it{Fe}})|$       &  44.675(37) \mbox{~T}                             &                       &  from the resonance frequency in line 4 and the \\
  ~ ~&                                              &                                           &                       &    magnetic moment value in line 3  \\
   \hline
       \\[-3mm]
     6  &    $|\nu(^{110}{\rm{Ag}}^{\textit{m}}Fe)|$               &  203.75(10) \mbox{~MHz}                        &  \cite{fox71}         &  NMR/ON, in $B_{\rm{ext}}$~=~0.220~T          \\
  7  &    $|\it{B}_{\rm{hf}}({\rm{Ag}}{\it{Fe}})|$       &  44.661(43) \mbox{~T}                             &                       &  from the resonance frequency in line 6 and the \\
  ~ ~&                                              &                                           &                       &    magnetic moment value in line 3  \\
   \hline
       \\[-3mm]
  8  &    $|\nu(^{106}{\rm{Ag}}^{\textit{m}}Fe)|$               &  210.57(3) \mbox{~MHz}                            &  \cite{eder85}        &  NMR/ON, for $B_{\rm{ext}}$~=~0~T             \\
  9  &    $|\nu(^{106}{\rm{Ag}}^{\textit{m}}Ag)|$               &  56.128(26) \mbox{~MHz}                           &  \cite{ohya01}        &  BF-NMR/ON~\footnotemark[3] in $B_{\rm{tot}}$~=~11.987~T          \\
 10  &    $|\it{B}_{\rm{hf}}({\rm{Ag}}{\it{Fe}})|$       &  44.720(43) \mbox{~T}                             &                       &  from the ratio of the resonance frequencies in lines 8 \\
  ~ ~&                                              &                                           &                       &    and 9 (see also Eq.~(\ref{eq:ratio_106mAg}))  \\
    \hline
       \\[-3mm]
 11  &    $|\nu(^{104}{\rm{Ag}}^{\textit{g}}Fe)|$               &  266.70(5) \mbox{~MHz}                            &  this work            &  NMR/ON, in $B_{\rm{ext}}$~=~0.1008~T          \\
 12  &    $|\mu(^{104}{\rm{Ag}}^{\textit{g}})|$                 &  3.919(3) \mbox{~$\mu_{\rm{N}}$}                  &  \cite{dinger89}      &  optical spectroscopy  \\
 13  &    $|\it{B}_{\rm{hf}}({\rm{Ag}}{\it{Fe}})|$       &  44.709(35) \mbox{~T}                             &                       &  from the resonance frequency of this work (line 11) \\
  ~ ~&                                              &                                           &                       &    and the magnetic moment value in line 12  \\
   \hline
   \hline
       \\[-3mm]
 14  &    $|\it{B}_{\rm{hf}}({\rm{Ag}}{\it{Fe}})|$  &  44.692(30) \mbox{~T}      &                       &  weighted average of the values in lines 5, 7, 10 and 13  \\

\end{tabular}
\end{ruledtabular}
\end{center}
\vspace*{-4mm}

\footnotetext[1] {Corrected for diamagnetism according to \cite{raghavan89}.}
\footnotetext[2] {Atomic beam magnetic resonance.}
\footnotetext[3] {Brute-force nuclear magnetic resonance on oriented nuclei.}
\end{table*}

\subsection{Nuclear magnetic moment of $^{104m}$Ag}

With the value for the hyperfine magnetic field of Ag impurities in
Fe being established we can now deduce a precise value for the
magnetic moment of $^{104}$Ag$^{\textit{m}}$ from our resonance frequency
$|\nu(^{104}{\rm{Ag}}^{\textit{m}}Fe)|$~=~627.7(4)~MHz. Taking into account the
external field $B_{\rm{ext}}$=0.1008(3)~T and the Knight shift
parameter $K$ = -0.046(5), and correcting for the demagnetization
field opposite to the direction of the external field, i.e.
$B_{\rm{dem}}$~=~0.026(5)~T  (see above), the total magnetic field
experienced by the nuclei is found to be $|B_{\rm{tot}}|$ = 44.622(30)~T.
Eq.~(\ref{eq:resFreqPractical}) then yields for the magnetic moment
\begin{linenomath*}
\begin{equation}
|\mu(^{104}{\rm{Ag}}^{\textit{m}})| = 3.691(3)~\mu_{\rm{N}}.
\end{equation}
\end{linenomath*}
\noindent Correcting for diamagnetism does not change this value. This is the first
precision value for the magnetic moment of this isomer.

\section{Magnetic moments of $^{102-110}$Ag}

The magnetic moments of the odd-odd even-A Ag isotopes are listed in
Table~\ref{tab:magMomentsExpMoments}.

%
\begin{table}
\begin{center}
\small \caption{Experimental magnetic moment values for the
ground and isomeric states of the even-A isotopes $^{102-110}$Ag.
Whenever a diamagnetic correction
was applied this is indicated. In all other cases the correction was either
negligibly small, or not necessary because of the method that was applied or because of
the large experimental error bar.}
\label{tab:magMomentsExpMoments} \smallskip
\begin{ruledtabular}
\begin{tabular*}{\columnwidth}{@{\extracolsep{\fill}}cc d l}
       \\[-3mm]
   Isotope & $I$  & \multicolumn{1}{c}{\mbox{\hspace*{8mm} $\mu$ ($\mu_{\rm{N}}$)}} & Ref. \\
       \\[-3mm]
  \hline
       \\[-3mm]
   $^{102}{\rm{Ag}}^{\textit{g}}$ & 5 &  4.55(65)                  & \cite{vandeplassche85}\footnotemark[1] \\
   $^{104}{\rm{Ag}}^{\textit{g}}$ & 5 &  3.916(8)\footnotemark[2]  & \cite{vandeplassche86,vanwalle85}\footnotemark[3]\\
               &   &  3.919(3)                  & \cite{dinger89}\footnotemark[4] \\
               &   & +4.0(2)                    & \cite{ames61}\footnotemark[5] \\
   $^{106}{\rm{Ag}}^{\textit{g}}$ & 1 & +2.83(20)                  & \cite{greenebaum74}\footnotemark[5] \\
   $^{108}{\rm{Ag}}^{\textit{g}}$ & 1 & 2.6884(7)\footnotemark[6]  & \cite{winnacker76}\footnotemark[3] \\
   $^{110}{\rm{Ag}}^{\textit{g}}$ & 1 & 2.7271(8)\footnotemark[6]  & \cite{winnacker76}\footnotemark[3] \\
       \\[-3mm]
   \hline
       \\[-3mm]
   $^{102}{\rm{Ag}}^{\textit{m}}$ & 2 & +4.12(25)                  & \cite{greenebaum74}\footnotemark[5] \\
   $^{104}{\rm{Ag}}^{\textit{m}}$ & 2 &  3.691(3)                  & this work\footnotemark[3] \\
               &   &  3.7(1)                    & this work\footnotemark[7] \\
               &   & +3.7(2)                    & \cite{ames61}\footnotemark[5] \\
   $^{106}{\rm{Ag}}^{\textit{m}}$ & 6 & 3.705(4)                   & \cite{ohya01}\footnotemark[8] \\
               &   & 3.709(4)                   & \cite{eder84}\footnotemark[3] \\
               &   & 3.71(15)                   & \cite{haroutunian76}\footnotemark[1] \\
               &   & 3.82(8)                    & \cite{berkes84}\footnotemark[1] \\
   $^{108}{\rm{Ag}}^{\textit{m}}$ & 6 & 3.577(20)                  & \cite{fischer75}\footnotemark[4] \\
   $^{110}{\rm{Ag}}^{\textit{m}}$ & 6 & 3.607(4)\footnotemark[2],\footnotemark[6]   & \cite{schmelling67}\footnotemark[5] \\
               &   & 3.609(4)\footnotemark[2],\footnotemark[6]   & \cite{hutchison92}\footnotemark[8] \\

\end{tabular*}
\end{ruledtabular}
\end{center}
\vspace*{-4mm}

\footnotetext[1]{Low Temperature Nuclear Orientation (LTNO).}
\footnotetext[2]{Recalculated using the hyperfine field value obtained in the previous section.}
\footnotetext[3]{Nuclear Magnetic Resonance on Oriented Nuclei (NMR/ON).}
\footnotetext[4]{Optical hyperfine spectroscopy.}
\footnotetext[5]{Atomic Beam Magnetic Resonance (ABMR).}
\footnotetext[6]{Corrected for diamagnetism.}
\footnotetext[7]{Resonant Laser Ionization Source technique (RILIS).}
\footnotetext[8]{Brute-Force Nuclear Magnetic Resonance on Oriented Nuclei (BF-NMR/ON).}
\end{table}

Extended magnetic moment calculations for even-A odd-odd Ag isotopes
are not available in the literature. Table~\ref{tab:Ag moments
overview} presents an overview of the experimental values for the
magnetic moments of the ground and isomeric states in the even-A
$^{102-110}$Ag isotopes as well as a comparison with values
calculated with the additivity rule~\cite{brennan60}
\begin{linenomath*}
\begin{equation}\label{eq:additivityRules}
    \mu_I [\mu_N] = \frac {1}{2} I (g_n + g_p) +
    \frac {
    (g_n - g_p)
    \left[
        I_n(I_n + 1) - I_p(I_p + 1)
    \right]
          }  {2 (I+1) }.
\end{equation}
\end{linenomath*}
For this we used Schmidt single particle moments, modified
single particle moments, as well as the mean
values of experimental magnetic moments of neighboring odd-even nuclei.

The Schmidt single particle magnetic moments for the $\pi
\rm{g}_{9/2}$ and $\nu \rm{d}_{5/2}$ orbitals that determine the
spins and magnetic moments of the light $^{101-110}$Ag nuclei are
+6.793~$\mu_{\rm{N}}$ and -1.913~$\mu_{\rm{N}}$, respectively
\cite{noya58}. Modified single particle moments take into account
the effects of the nuclear medium, i.e. configuration mixing, core
polarization and meson exchange \cite{kumar75} by using the
effective gyromagnetic ratios $g_l^{\rm{mod}}(\pi)$ = +1.1,
$g_l^{\rm{mod}}(\nu)$ = -0.05 and $g_s^{\rm{mod}}$ = 0.7
$g_s^{\rm{free}}$, i.e. $g_s^{\rm{mod}}(\pi)$ = 3.910 and
$g_s^{\rm{mod}}(\nu)$ = -2.678, thus yielding
$\mu^{\rm{mod}}(\pi(\rm{g}_{9/2}))$ = +6.355~$\mu_{\rm{N}}$ and
$\mu^{\rm{mod}}(\nu(\rm{d}_{5/2}))$ = -1.439 $\mu_{\rm{N}}$.
Finally, using experimental $g$~factors from neighboring odd-even
nuclei has the advantage that configuration mixing and possible $g$
factor quenching in the odd-$A$ nuclei are automatically taken into
account. For the $g$ factor of the $\rm{g}_{9/2}$ protons, the value
of the lower mass odd-A silver isotope was each time used. For the
$g$ factor of the $\rm{d}_{5/2}$ neutrons the geometrical mean of
the $g$ factors of the neighboring isotopes of Ru, Pd and Cd with
the neutron in the $\rm{d}_{5/2}$ orbital were used.

Note that nearly all neighboring odd-neutron Ru, Pd, Cd and Sn isotopes
with N = 55, 57, 59, 61 and 63 have a 5/2$^+$ ground state. For the
isotopes with N = 55 this is due to the hole in the $\nu \rm{d}_{5/2}$
orbit. For the isotopes with N = 57 to 63, filling the
$\nu \rm{g}_{7/2}$ orbit, this indicates that with
adding neutron pairs these predominantly couple to spin 0, leaving
the hole in the $\nu \rm{d}_{5/2}$ orbit, and this all the way up to N = 64.
This is why in the calculations of magnetic moments presented in
Table~\ref{tab:Ag moments overview} the $(\nu \rm{d}_{5/2})^{-1}$
configuration has always been used.
The 7/2$^+$ state related to a filled $\nu \rm{d}_{5/2}$ orbit
and an odd number of neutrons in the $\nu \rm{g}_{7/2}$ orbit
is in almost all isotopes found as an excited state
at excitation energies ranging from 188~keV to 416~keV, except for
$^{111}$Sn where it is the ground state. It is not understood why
this predominant coupling of neutron pairs in the $\nu \rm{g}_{7/2}$
orbit occurs. However, the configuration is complex as can be seen
from the g factors of the odd-neutron states in this mass
region which are found to have values of about -0.30 (see column 5 in
Table~\ref{tab:Ag moments overview}) which in absolute value is
considerably smaller than the modified Schmidt value of -0.576
for the $\nu \rm{d}_{5/2}$ orbit (the modified Schmidt g factor
for the $\nu \rm{g}_{7/2}$ orbit is +0.242).
Note that this issue is to some extent also related
to the current debate about the ground state of $^{101}$Sn being
$\nu \rm{d}_{5/2}$ or $\nu \rm{g}_{7/2}$ (e.g.
\cite{kavatsyuk07,seweryniak09}).

As can be seen in Table~\ref{tab:Ag moments overview}, for the
heavier neutron deficient isotopes $^{106}$Ag, $^{108}$Ag, and
$^{110}$Ag (with a $1^+$ ground state and a $6^+$ isomeric state)
the magnetic moment for the isomeric state can be very well
explained by a parallel coupled $(\pi \rm{g}_{9/2})^{-3}_{7/2^+}
(\nu \rm{d}_{5/2})^{-1}_{5/2^+}$ configuration. The values
calculated with the additivity relation
(Eq.~(\ref{eq:additivityRules})) using Schmidt single particle
moments or modified single particle moments are systematically lower
than the experimental values, but the values calculated with
experimental $g$~factors are always very close to the experimental
results.
%

For the 1$^+$ ground state of $^{106, 108, 110}$Ag, coming from the
antiparallel coupling of the $(\pi \rm{g}_{9/2})^{-3}_{7/2^+}$
proton group and the $(\nu \rm{d}_{5/2})^{-1}_{5/2^+}$ neutron
state, the additivity rule with experimental $g$~factors of
neighboring isotopes yields values that are systematically about 0.3
to 0.5~$\mu_{\rm{N}}$ too large. Since the correction on this 1$^+$
state due to core excitations is less than 0.1~$\mu_{\rm{N}}$
\cite{noya58} this deviation is probably due to the mixing of other
states in the wave function. The $(\pi \rm{g}_{9/2})^{-3}_{{7/2}^+}$
proton configuration is thus firmly established for $^{106-110}$Ag.

\begin{table*}[here]
\begin{center}
\small \caption{Magnetic moments of the even-A Ag isotopes.
Experimental magnetic moments are compared with values
obtained with the addition theorem (Eq.~(\ref{eq:additivityRules}))
using either Schmidt values, $\mu_{\rm{add}}^{\rm{Sch}}$,
modified Schmidt values, $\mu_{\rm{add}}^{\rm{mod}}$,
or experimental moment values from neighboring isotopes,
$\mu_{\rm{add}}^{\rm{exp}}$. Note that in fact the $(\nu \rm{d}_{5/2})^{-1}$
neutron configuration was used to calculate $\mu_{\rm{add}}^{\rm{Sch}}$
and $\mu_{\rm{add}}^{\rm{mod}}$ (see text). Since $g_{n,\rm{exp}}$ was obtained from
neighbouring isotopes with $I^{\pi}$ = 5/2$^+$, the values for
$\mu_{\rm{add}}^{\rm{exp}}$ in column 8 take into account the fact
that the effective neutron configuration is a mixture of
$\nu \rm{d}_{5/2}$ and $\nu \rm{g}_{7/2}$ coupling to a spin 5/2$^+$.
Further, since the Schmidt g-factors for $\nu \rm{d}_{5/2}$ and
$\nu \rm{g}_{7/2}$ are -0.765 and 0.425, respectively, the fact that
$g_{n,\rm{exp}}$ varies between -0.252 and -0.0331 (see column 5)
indicates that a considerable amount of mixing between the
$\nu \rm{d}_{5/2}$ and $\nu \rm{g}_{7/2}$ states to be present.
(table adapted from Ref.~\cite{vanwalle85})}
\label{tab:Ag moments overview}
\smallskip
\begin{ruledtabular}
\begin{tabular} {ccccc cccc cc}
       \\[-3mm]
  $A$ & $I^{\pi}$ & configuration &
         $g_{p,\rm{exp}}$
      & $g_{n,\rm{exp}}$
      & \begin{tabular}{c}
          $\mu_{\rm{add}}^{\rm{Sch}}$ \\
          $\mu_{\rm{N}}$ \\
        \end{tabular}
      & \begin{tabular}{c}
          $\mu_{\rm{add}}^{\rm{mod}}$ \\
          $\mu_{\rm{N}}$ \\
        \end{tabular}
      & \begin{tabular}{c}
          $\mu_{\rm{add}}^{\rm{exp}}$ \\
          $\mu_{\rm{N}}$ \\
        \end{tabular}
      & \begin{tabular}{c}
          $\mu_{\rm{exp}}$ \\
          $\mu_{\rm{N}}$ \\
        \end{tabular}
      & \begin{tabular}{c}
          Ref. \\
          $\mu_{\rm{exp}}$ \\
        \end{tabular}
      &  \begin{tabular}{c}
          mix. \\
          ampl. \\
          $\alpha$ or $\beta$ \\
          (Eq.(~\ref{eq:expectationMagneMoment})) \\
         \end{tabular}
       \\
       \\[-3mm]
\hline
       \\[-3mm]
  102 & $2^{+}$
      & $\left[  \left(\pi \rm{g}_{9/2} \right)_{7/2^{+}}^{-3} \left( \nu \rm{d}_{5/2} ~ \nu \rm{g}_{7/2} \right)_{5/2^{+}}
         \right]_{2^{+}}$
      & $1.250(2)$\footnotemark[1] & $-0.279(29)$\footnotemark[2]  & 3.40 & 3.16 & 2.76(1) & 4.12(25) & \cite{greenebaum74} & 0.66(11) \\
       \\[-3mm]
      &    & $\left[  \left(\pi \rm{g}_{9/2} \right)_{9/2^{+}}^{-3} \left( \nu \rm{d}_{5/2} ~ \nu \rm{g}_{7/2} \right)_{5/2^{+}}
              \right]_{2^{+}}$
      &  &  & 6.81 & 6.14 & 5.05(5) &  &  & 0.75(9) \\
       \\[-3mm]
      & $5^{+}$  & $\left[  \left(\pi \rm{g}_{9/2} \right)_{7/2^{+}}^{-3} \left( \nu \rm{d}_{5/2} ~ \nu \rm{g}_{7/2} \right)_{5/2^{+}}
         \right]_{5^{+}}$
      &  &  & 3.19 & 3.25 & 3.32(6) & 4.55(65) & \cite{vandeplassche85} & 0.0(3) \\
       \\[-3mm]
      &    & $\left[  \left(\pi \rm{g}_{9/2} \right)_{9/2^{+}}^{-3} \left( \nu \rm{d}_{5/2} ~ \nu \rm{g}_{7/2} \right)_{5/2^{+}}
              \right]_{5^{+}}$
      &   &   & 4.90 & 4.74 & 4.47(3) &  &  & 1.0(3) \\
       \\[-3mm]
\hline
       \\[-3mm]
  104 & $2^{+}$
      & $\left[  \left(\pi \rm{g}_{9/2} \right)_{7/2^{+}}^{-3} \left( \nu \rm{d}_{5/2} ~ \nu \rm{g}_{7/2} \right)_{5/2^{+}}
         \right]_{2^{+}}$
      & $1.266(1)$\footnotemark[3] & $-0.294(4)$\footnotemark[4]  & 3.40 & 3.16 & 2.792(2) & 3.691(3)\footnotemark[5] & this work & 0.79(1) \\
       \\[-3mm]
      &    & $\left[  \left(\pi \rm{g}_{9/2} \right)_{9/2^{+}}^{-3} \left( \nu \rm{d}_{5/2} ~ \nu \rm{g}_{7/2} \right)_{5/2^{+}}
              \right]_{2^{+}}$
      &  &  & 6.81 & 6.14 & 5.129(5) &  &  & 0.61(2) \\
       \\[-3mm]
      & $5^{+}$  & $\left[  \left(\pi \rm{g}_{9/2} \right)_{7/2^{+}}^{-3} \left( \nu \rm{d}_{5/2} ~ \nu \rm{g}_{7/2} \right)_{5/2^{+}}
         \right]_{5^{+}}$
      &  &  & 3.19 & 3.25 & 3.356(8) & 3.919(3) \footnotemark[6] & \cite{dinger89} & 0.72(2) \\
       \\[-3mm]
      &    & $\left[  \left(\pi \rm{g}_{9/2} \right)_{9/2^{+}}^{-3} \left( \nu \rm{d}_{5/2} ~ \nu \rm{g}_{7/2} \right)_{5/2^{+}}
              \right]_{5^{+}}$
      &  &   & 4.90 & 4.74 & 4.525(6) &  &  & 0.69(2) \\
       \\[-3mm]
\hline
       \\[-3mm]
  106 & $1^{+}$
      & $\left[  \left(\pi \rm{g}_{9/2} \right)_{7/2^{+}}^{-3} \left( \nu \rm{d}_{5/2} ~ \nu \rm{g}_{7/2} \right)_{5/2^{+}}
         \right]_{1^{+}}$
      & $1.261(4)$\footnotemark[7] & $-0.252(6)$\footnotemark[8]  & 4.35 & 3.90 & 3.15(2) & 2.83(20) & \cite{greenebaum74} & -- \\
       \\[-3mm]
      &   &   &   &  &   &   &   &   &   &   \\
       \\[-3mm]
      & $6^{+}$
      & $\left[  \left(\pi \rm{g}_{9/2} \right)_{7/2^{+}}^{-3} \left( \nu \rm{d}_{5/2} ~ \nu \rm{g}_{7/2} \right)_{5/2^{+}}
         \right]_{6^{+}}$
      &  &   & 3.38 & 3.50 & 3.79(2) & 3.705(4)\footnotemark[9]  & \cite{ohya01} & -- \\
       \\[-3mm]
      &   &   &   &   &   &   &   &   &   &   \\
       \\[-3mm]
\hline
       \\[-3mm]
  108 & $1^{+}$
      & $\left[  \left(\pi \rm{g}_{9/2} \right)_{7/2^{+}}^{-3} \left( \nu \rm{d}_{5/2} ~ \nu \rm{g}_{7/2} \right)_{5/2^{+}}
         \right]_{1^{+}}$
      & $1.257(2)$\footnotemark[10] & $-0.3311384(6)$\footnotemark[11]  & 4.35 & 3.90 & 3.242(5) & 2.6884(7) & \cite{winnacker76} & -- \\
       \\[-3mm]
      &   &   &   &  &   &   &   &   &   &   \\
       \\[-3mm]
      & $6^{+}$
      & $\left[  \left(\pi \rm{g}_{9/2} \right)_{7/2^{+}}^{-3} \left( \nu \rm{d}_{5/2} ~ \nu \rm{g}_{7/2} \right)_{5/2^{+}}
         \right]_{6^{+}}$
      &  &   & 3.37 & 3.50 & 3.572(7) & 3.577(20) & \cite{fischer75} & -- \\
       \\[-3mm]
      &   &   &   &   &   &   &   &   &   &   \\
       \\[-3mm]
\hline
       \\[-3mm]
  110 & $1^{+}$
      & $\left[  \left(\pi \rm{g}_{9/2} \right)_{7/2^{+}}^{-3} \left( \nu \rm{d}_{5/2} ~ \nu \rm{g}_{7/2} \right)_{5/2^{+}}
         \right]_{1^{+}}$
      & $1.257(2)$\footnotemark[12] & $-0.306(1)$\footnotemark[13]  & 4.35 & 3.90 & 3.211(5) & 2.7271(8) & \cite{winnacker76} & -- \\
       \\[-3mm]
      &   &   &   &  &   &   &   &   &   &   \\
       \\[-3mm]
      & $6^{+}$
      & $\left[  \left(\pi \rm{g}_{9/2} \right)_{7/2^{+}}^{-3} \left( \nu \rm{d}_{5/2} ~ \nu \rm{g}_{7/2} \right)_{5/2^{+}}
         \right]_{6^{+}}$
      &  &   & 3.37 & 3.50 & 3.635(7) & 3.609(4)\footnotemark[14] & \cite{hutchison92} & -- \\
       \\[-3mm]
      &   &   &   &   &   &   &   &    &    &   \\
       \\[-3mm]
\end{tabular}
\end{ruledtabular}
\end{center}

\begin{minipage}[b]{1.\linewidth}
\vspace*{-4mm}

\footnotetext[1]{$g$ factor of $^{101}$Ag (I$^\pi$ = 9/2$^+$) from
Ref.~\cite{dinger89}.}

\footnotetext[2]{Average of the $g$ factors
of $^{99}$Ru, $^{101}$Pd and $^{103}$Cd.}

\footnotetext[3]{$g$
factor of $^{103}$Ag (I$^\pi$ = 7/2$^+$) from Ref.~\cite{dinger89}.}

\footnotetext[4]{Average of the $g$ factors of $^{101}$Ru and
$^{105}$Cd.}

\footnotetext[5]{Other determinations of this magnetic
moment have yielded +3.7(2)~$\mu_{\rm{N}}$~\cite{ames61} and
3.7(1)~$\mu_{\rm{N}}$ (this work, RILIS measurement).}

\footnotetext[6]{Other determinations of this magnetic moment have
yielded 3.916(8)~$\mu_{\rm{N}}$~\cite{vandeplassche86}
(see also Table~\ref{tab:magMomentsExpMoments}), and
+4.0(2)~$\mu_{\rm{N}}$~\cite{ames61}.}

\footnotetext[7]{$g$ factor
of $^{105}$Ag$^{\textit{m}}$ (I$^\pi$ = 7/2$^+$) from Ref.~\cite{dinger89}.}

\footnotetext[8]{Average of the $g$ factors of $^{105}$Pd and
$^{107}$Cd.}

\footnotetext[9]{Other determinations of this magnetic
moment have yielded 3.709(4)~$\mu_{\rm{N}}$~\cite{eder84},
3.71(15)~$\mu_{\rm{N}}$~\cite{haroutunian76}, and
3.82(8)~$\mu_{\rm{N}}$~\cite{berkes84}.}

\footnotetext[10]{$g$
factor of $^{107}$Ag$^{\textit{m}}$ (I$^\pi$ = 7/2$^+$) from Ref.~\cite{eder85}.}

\footnotetext[11]{$g$ factor of $^{109}$Cd from
Ref.~\cite{spence72}.}

\footnotetext[12]{$g$ factor of $^{109}$Ag$^{\textit{m}}$
(I$^\pi$ = 7/2$^+$) from Ref.~\cite{eder85}.}

\footnotetext[13]{$g$ factor of $^{111}$Cd (level at 245 keV) from Ref.~\cite{bertschat74}.}

\footnotetext[14]{Another determination of this magnetic moment has
yielded 3.607(4)~$\mu_{\rm{N}}$~\cite{schmelling67}.}

\end{minipage}%
\end{table*}
%

%

For the lower mass $^{102,104}$Ag isotopes Ames et al.~\cite{ames61}
already pointed out that the wave function would consist of a $(\nu
\rm{d}_{5/2})^{-1}_{5/2^+}$ neutron configuration coupled to a mixed
$(\pi\rm{g}_{9/2})^{-3}_{9/2^+}$- $(\pi \rm{g}_{9/2})^{-3}_{7/2^+}$
proton configuration. The experimental magnetic moments of both the
ground and the isomeric state in $^{104}$Ag are in between the
values calculated for the two pure proton-neutron configurations
(see Table ~\ref{tab:Ag moments overview}) and support this
suggestion. The magnetic moment of $^{104}$Ag$^{\textit{g}}$ ($I^\pi$ = 5$^+$) was
previously already determined with high precision in two independent
experiments \cite{vandeplassche86,dinger89}. The measurement
reported here now provides a precise value for $^{104}$Ag$^{\textit{m}}$ ($I^\pi$
= 2$^+$) as well. This now permits a more detailed analysis of the
configuration mixture in the $2^+, 5^+$ doublet of states (only
6.9~keV apart) in $^{104}$Ag. Writing the wave functions for both
states as (see \cite{noya58}):
\begin{linenomath*}
\begin{eqnarray}\label{eq:waveFunction}
    \psi(^{104}\rm{Ag}^{\textit{g,m}}) =&& \alpha
    \left[
        (\pi \rm{g}_{9/2})^{-3}_{{7/2}^+} (\nu \rm{d}_{5/2})^{-1}_{{5/2}^+}
    \right] + \nonumber\\
    && \beta
    \left[
        (\pi \rm{g}_{9/2})^{-3}_{{9/2}^+} (\nu \rm{d}_{5/2})^{-1}_{{5/2}^+}
    \right]  ,
\end{eqnarray}
\end{linenomath*}
\noindent with $\alpha$ and $\beta = \sqrt{1-\alpha^2}$ the mixing amplitudes,
the expectation value of the magnetic moment operator becomes
\begin{linenomath*}
\begin{eqnarray}\label{eq:expectationMagneMoment}
    <\mu> = && \alpha^2
    \left<
         \mu \left[ (\pi \rm{g}_{9/2})^{-3}_{{7/2}^+} (\nu \rm{d}_{5/2})^{-1}_{{5/2}^+} \right]
    \right> + \nonumber\\
    && \beta^2
        \left<
         \mu \left[ (\pi \rm{g}_{9/2})^{-3}_{{9/2}^+} \nu (d_{5/2})^{-1}_{{5/2}^+} \right]
    \right>    .
\end{eqnarray}
\end{linenomath*}
The mixing amplitudes, as
obtained from the experimental magnetic moments, are listed in the last column
of Table~\ref{tab:Ag moments overview} and indicate a complete mixing
of the two configurations (i.e. $\alpha \approx \beta \approx 1/\sqrt{2}$ =
0.71) in $^{104}$Ag$^{\textit{g}}$ ($I^\pi=5^+$) and almost complete mixing in
$^{104}$Ag$^{\textit{m}}$ ($I^\pi=2^+$).
Performing the same analysis also for the doublet of 2$^+$ and 5$^+$
states of $^{102}$Ag$^{\textit{m}}$ and $^{102}$Ag$^{\textit{g}}$, respectively (which differ
only 9.3~keV in energy), indicates also almost complete mixing of the
two different proton-neutron configurations in $^{102}$Ag$^{\textit{m}}$ ($I^\pi$
= 2$^+$), although the error bars still allow for a smaller contribution
from the $(\pi \rm{g}_{9/2})^{-3}_{{7/2}^+}$ proton configuration. As for
$^{102}$Ag$^{\textit{g}}$ ($I^\pi=5^{+}$), the experimental magnetic moment value
suggests an almost pure $(\pi \rm{g}_{9/2})^{-3}_{{9/2}^+}$
configuration, although the large error bar still allows for a
sizeable mixing from the $(\pi \rm{g}_{9/2})^{-3}_{{7/2}^+}$
configuration (see Table~\ref{tab:Ag moments overview}). A larger
contribution of the $(\pi \rm{g}_{9/2})^{-3}_{{9/2}^+}$
configuration in $^{102}$Ag$^{\textit{g}}$, as is apparently observed,
would be in line with the fact that the ground state
in $^{101}$Ag has $I^\pi$ = 9/2$^+$ and also with the magnetic moment of
$^{101}$Ag. Indeed, the experimental magnetic moment of $^{101}$Ag,
i.e. $\mu$ = 5.627(11)~$\mu_{\rm{N}}$ \cite{dinger89}, is very close
to the value of 5.67~$\mu_{\rm{N}}$ that was calculated in
Ref.~\cite{paar73} for the lowest 9/2$^+$ state in the odd Ag
isotopes based on a configuration that is dominated by the $(\pi
\rm{g}_{9/2})^{-3}_{{9/2}^+}$ proton group.
New and precise measurements of the magnetic moments of the ground and
isomeric state of $^{102}$Ag as well as of the lower mass even-A Ag
isotopes could help to further clarify this. An interesting new method
in this respect, viz. in-gas-cell laser spectroscopy \cite{cocolios09},
was recently reported. This method is currently being applied to the
isotopes $^{97-102}$Ag \cite{darby10}.

\section*{ACKNOWLEDGEMENTS}

We would like to thank the technical staff of the ISOLDE
collaboration at CERN, Switzerland.
This work was supported
by FWO-Vlaanderen (Belgium), GOA/2004/03 (BOF-K.U.Leuven),
the 'Interuniversity Attraction Poles Programme–
Belgian State-Belgian Science Policy' (BriX network P6/23),
the European Commission within the Sixth Framework
Programme through I3-EURONS (Contract RII3-CT-2004-
506065), and the Grants of the Ministry of Education of the
Czech Republic 1P04LA211 and LA08015.



\end{document}